%% file: paper.tex
\documentclass[12pt]{article}
\usepackage{epsfig}
\usepackage{latexsym}

\newif\iffigure
\figuretrue

\newif\iftable 
\tabletrue

\newif\ifFINAL
\FINALtrue               %% UPDATED version of paper

\let\onlinecite=\cite
\input{rgmac.latex}

\def\gsim{\mathrel{\raise2pt\hbox to 8pt{\raise -5pt\hbox{$\sim$}\hss{$>$}}}}
\def\rsim{\mathrel{\raise2pt\hbox to 8pt{\raise -5pt\hbox{$\sim$}\hss{$>$}}}}
\def\lsim{\mathrel{\raise2pt\hbox to 8pt{\raise -5pt\hbox{$\sim$}\hss{$<$}}}}

\def\summ{\hbox{$m_{u}+m_{d}\ $}}
\def\summpunc{$m_{u}+m_{d}$}
\def\mbar{\hbox{$m_{u}+m_{d}$}}
\def\sdual{$s_0\ $}
\def\msbar{$\bar{MS}\ $}
\def\specf{$\rho_5(t)\ $}
\def\specfpunc{$\rho_5(t)$}
\def\specfs{$\rho_5(s)\ $}
\def\specfspunc{$\rho_5(s)$}
\def\piprime{$\pi^\prime (1300)\ $}
\def\piprprime{$\pi^{\prime\prime}(1800)$}
\def\me{$\langle 0\vert\partial_\mu A^\mu\vert 3\pi\rangle$}
\def\strangem{$m_s$}
\newcommand{\be}{\begin{equation}}
\newcommand{\ee}{\end{equation}}
\newcommand{\bea}{\begin{eqnarray}}
\newcommand{\eea}{\end{eqnarray}}

\newcommand\figcaption[1]{\vskip-0.0truein\caption{#1}\vskip0.0truein}

\newcommand\MSbar{\hbox{$\overline{MS}$}}

\def\cpt{\hbox{$ \chi $PT}}

\def\vev\#1{\langle \#1 \rangle}

\begin{document}

\begin{titlepage}
 \null
 \begin{center}
 \makebox[\textwidth][r]{LAUR-96-2698}
%%% \makebox[\textwidth][r]{40614-43}
 \par\vspace{0.25in} %%%
  {\Large
The Extraction of Light Quark Masses From Sum Rule Analyses \\
of Axial and Vector Current Ward Identities}
  \par
 \vskip 2.0em
 
 {\large 
  \begin{tabular}[t]{c}
        Tanmoy Bhattacharya \footnotemark\ and
         Rajan Gupta \footnotemark\\[0.5em]
        \em Group T-8, Mail Stop B-285, Los Alamos National Laboratory\\
        \em Los Alamos, NM 87545, U.~S.~A\\[1.5em]
        Kim Maltman \footnotemark\\[0.5em]
        \em Department of Mathematics and Statistics, York University \\
        \em 4700 Keele Street, North York, Ontario, Canada M3J1P3 \\
	\em and \\
	\em Special Research Centre for the Subatomic Structure of Matter \\
	\em University of Adelaide, Australia 5005 \\
  \end{tabular}}
 \par \vskip 2.0em
 {\large\bf Abstract}
\end{center}

\quotation In light of recent lattice results for the light quark
masses, $m_s$ and $m_u+m_d$, we re-examine the use of sum rules in the
extraction of these quantities, and discuss a number of potential
problems with existing analyses.  The most important issue is that of
the overall normalization of the hadronic spectral functions relevant
to the sum rule analyses. 
We explain why previous treatments, which fix this normalization by
assuming complete resonance dominance of the continuum threshold
region can potentially overestimate the resonance contributions
to spectral integrals by factors as large as $\sim 5$.
%We claim that the normalization should be
%fixed by the value at resonance, and {\it not} by that at threshold. We
%explain why previous treatments which fix the normalization by
%assuming that the threshold values are completely saturated by
%resonance contributions overestimate the spectral integrals by as much as 
%a factor of $4$.
We propose an alternate method of normalization
based on an understanding of the role of resonances in Chiral
Perturbation Theory which avoids this problem.  
The second important uncertainty we consider
relates to the physical content of the assumed location, $s_0$, of
the onset of duality with perturbative QCD.  We find that the
extracted quark masses depend very sensitively on this parameter.  We
show that the assumption of duality imposes very severe
constraints on the shape of the relevant spectral function in the dual
region, and present rigorous lower bounds for $m_{u}+m_{d}$ as a
function of $s_0$ based on a combination of these constraints and
the requirement of positivity of $\rho_5(s)$.  In the extractions of
$m_s$ we find that the conventional choice of the value of $s_0$ is
not physical. For a more reasonable choice of $s_0$ we are not able to
find a solution that is stable with respect to variations of the Borel
transform parameter.  This problem can, unfortunately,  be overcome only
if the hadronic spectral function is determined up to significantly
larger values of $s$ than is currently possible. Finally, we also
estimate the error associated with the convergence of perturbative QCD
expressions used in the sum rule analyses.  Our conclusion is that,
taking all of these issues into account, the resulting sum rule
estimates for both $m_u+m_d$ and $m_s$ could easily
have uncertainties as large as 
%be lowered by 
a factor of $2$, which would
make them compatible with the low estimates obtained from lattice QCD.

% The following is a hack ... needed because hyperref
% does not understand what is needed
{
 \def\test{Hfootnote.3}
 \expandafter\ifx\csname @currentHref\endcsname\test
% Now we know we are in the correct version of hyperref
 \expandafter\def\csname @currentHref\endcsname{Hfootnote.1}
 \footnotetext[1]{Email: tanmoy@qcd.lanl.gov}
 \expandafter\def\csname @currentHref\endcsname{Hfootnote.2}
 \footnotetext[2]{Email:    rajan@qcd.lanl.gov}
 \expandafter\def\csname @currentHref\endcsname{Hfootnote.3}
 \footnotetext[3]{Email:    maltman@fewbody.phys.yorku.ca}
 \else
 \footnotetext[1]{Email: tanmoy@qcd.lanl.gov}
 \footnotetext[2]{Email:    rajan@qcd.lanl.gov}
 \footnotetext[3]{Email:    fs300175@sol.yorku.ca}
 \fi
}
\vfill
\mbox{24 MARCH, 1997}
\end{titlepage}

\setlength{\textfloatsep}{12pt plus 2pt minus 2pt}

\makeatletter % Thus allowing override of non-user parameters

% We do not want extra spacing between items
\setlength{\leftmargini}{\parindent}
\def\@listi{\leftmargin\leftmargini
            \topsep 0\p@ plus2\p@ minus2\p@\parsep 0\p@ plus\p@ minus\p@
            \itemsep \parsep}
% Do not want line break after table number in table captions.
\long\def\@maketablecaption#1#2{#1. #2\par}

% Allow a tiny stretch between paragraphs
\advance \parskip by 0pt plus 1pt minus 0pt

\makeatother

\section{INTRODUCTION}
\label{s:s_intro}

The recent lattice results for the light quark masses\cite{LANLmq96},
$\mbar=6.8 \pm 0.8 \pm 0.6 \MeV$ and $m_s = 100 \pm 21 \pm 10 \MeV$ in
the quenched approximation, and the even smaller values $\mbar = 5.4
\pm 0.6 \pm 0.6 \MeV$ and $m_s = 68 \pm 12 \pm 7 \MeV$, for the
$n_f=2$ flavor theory, (all evaluated in the \MSbar\ scheme at $\mu=2
\GeV$), appear to be significantly smaller than results obtained from
sum rule analyses.  The most recent and complete sum rules analyses
are those of ($i$) Bijnens, Prades and deRafael (BPR), which yields
$\mbar(\mu=2\GeV) = 9.4 \pm 1.76 \MeV$ \onlinecite{BPR95}, and ($ii$)
Chetyrkin, Pirjol, and Schilcher (CPS) which gives $m_s = 143 \pm 14
\MeV$ \onlinecite{CPS96}.  We have translated the original values,
$\mbar=12 \pm 2.5 \MeV$ and $m_s = 203 \pm 20 \MeV$, quoted at $\mu=1
\GeV$, to $\mu = 2 \GeV$ using the renormalization group running and
the preferred value $\Lambda_{QCD}^{(3)} = 300,\ 380 $ MeV used
respectively in the two calculations.  (The analysis by CPS is an
update of that by Jamin and M\"unz (JM) \onlinecite{jm95}, however,
since the approach and techniques are the same, we will refer to their
work jointly by the abbreviation JM/CPS.)  The sum rule results, thus,
lie roughly $1-2 \sigma$ above the quenched results.  The difference
between the sum rule and the $n_f=2$ lattice estimates, however, is
large and, we feel, significant enough to warrant scrutiny.  
\ifFINAL
Both the lattice and sum rules approaches have their share of
systematic errors. A recent review of the lattice results is given in \cite{Mq97TBRG}. 
Here we present a re-evaluation of the sum rules analyses. 
\else
Since both the lattice and sum rules approaches have their share of
systematic errors, we present a re-evaluation of both analyses in
order to explore the question of whether the remaining differences
should be considered significant or not.
\fi

The issues in the sum-rule analyses that we shall concentrate on are the
convergence of perturbative QCD (pQCD) expressions, the choice of
$s_0$ -- the scale beyond which quark-hadron duality is 
assumed to be valid, and the normalization of resonance contributions in
the ansatz for the hadronic spectral function for $s \le s_0$.  

The first issue is important because both the $\alpha_s$ and
$\alpha_s^2$ corrections to the 2-point correlation functions used in
sum rule analyses are large.  This issue has been analyzed in detail
by CPS for the extraction of $m_s$; therefore, we shall only comment on
it briefly for the case of $\mbar$.

The second point is important because, as we will show below, it turns
out that the extraction of the quark masses, in particular that of
\mbar, is very sensitive to the choice of \sdual. This is illustrated
by deriving lower bounds on \mbar\ associated with the positivity of
\specfpunc, and by investigating trial spectral functions.  Ideally
one would like to pick $s_0$ large enough so that pQCD, to the order
considered, can be shown to be reliable. Unfortunately, for larger
$s_0$, the hadronic spectral function receives contributions from an
increasing number of intermediate states, and hence becomes
increasingly hard to model.  We discuss the uncertainties introduced
by a compromise choice of $s_0$. In the extraction of $m_s$ by JM/CPS
we argue that an artificially large value of $s_0$ has been used. For
a more reasonable value of $s_0$ we are not able to find an estimate
for $m_s$ that is stable under variations of the Borel transform scale
$u$.

The third issue arises because the continuum part of the hadronic
spectral function is typically represented as a
sum-of-resonances-modulation of a continuum form, the overall
normalization of which is fixed by assuming complete resonance dominance of the
spectral function near continuum threshold.  This turns out to be
potentially the most important issue. We, in fact, show in the case
of the vector 2-point function, for which experimental information on
the spectral function is available in the resonance region, that an
analogous extrapolation from threshold to the $\rho$ meson peak would
lead to an overestimate of the spectral function in the resonance
region by a factor of $\sim 5$.  We then explain the origin of this
problem from the point of view of the existing phenomenological
understanding of how resonance contributions enter the expressions
for low-energy observables as computed in Chiral Perturbation Theory ($\cpt$).
Based on this understanding, we propose an alternate method for
normalizing the spectral function in the resonance region which
requires as input only the expression obtained from $\cpt$ to 1-loop
order, in the near-threshold region.  We 
then employ this method in a re-analysis of the only sum rule treatment for
which the relevant $\cpt$ expression is known, namely that of the
correlator of the product of two divergences of the
strangeness-changing vector current (as used by JM/CPS to obtain the
estimate quoted above for $m_s$) and show that the traditional method
of normalization leads to a significant
overestimate of $m_s$.

We find that the size of the corrections suggested by our
consideration of the above issues can easily lower the sum rule
estimates for both \mbar\ and $m_s$ by a factor at least as large
as two.  In particular, using the corrected normalization for the
hadronic spectral function in the JM/CPS analysis alone would lower
the extracted value of $m_s$ by almost exactly a factor of $2$.  Such
a change would make the lattice and sum rule estimates consistent.
Lowering both estimates by roughly the same factor would, moreover,
preserve agreement of the ratio, $r= 2m_s/(m_u+m_d)$, with that
predicted by $\cpt$.  
\ifFINAL
\else
For completeness, we also discuss the systematic
errors in lattice calculations.
\fi

The paper is organized as follows.  In order to make it
self-sufficient, and to introduce notation, we reproduce the necessary
details from Refs.~\onlinecite{BPR95} and \onlinecite{jm95} in
Sections~\ref{s:s_fser} and \ref{s:s_JM}.  The convergence of pQCD is
discussed in Section~\ref{s:sPQCD}.  In Section~\ref{s:spositive} we
derive lower bounds on \mbar, as a function of $s_0$, using the
positivity of the relevant spectral function, $\rho_5$.  In
Section~\ref{s:sguess} we illustrate the potential sensitivity of the
extracted value of \mbar\ to the choice of $s_0$ by considering a
number of plausible trial spectral functions.  The important issue of
the overall normalization of the hadronic spectral function is
investigated in Section~\ref{s:s_vector} using the vector current case
as an illustrative example.  Based on the lessons learned from the
vector channel, a re-analysis of the JM/CPS estimate of $m_s$ is
presented in section~\ref{s:s_JM}.  
\ifFINAL
\else
A brief discussion of the
systematic errors in lattice calculations is given in
Section~\ref{s:s_LQCD}. 
\fi
Finally we end with some conclusions in
Section~\ref{s:s_conclusions}.

\section{FINITE ENERGY SUM RULES}
\label{s:s_fser}
The standard starting point for the extraction of the light quark mass
combination, \summpunc, is the Ward identity relating the divergence
of the axial current to the pseudoscalar density,
\be 
\label{eq: div. axial current}
\partial^{\mu}A_{\mu}^{(\pm)}(x)=(m_{d}+m_{u})
  \bar{q}(x)i\gamma_{5}\frac{\lambda_{1} \pm i\lambda_{2}}{2}q(x)
\ee
where $\bar{q}\equiv (\bar{u},\bar{d},\bar{s})$, and the projections $\pm \equiv
(\lambda_{1}\pm i\lambda_2)/2$ pick out states with the quantum
numbers of the $\pi^\pm$. 
This relation implies, for the two-point function of the
product of two such divergences, that 
\bea
\label{eq: two-point funct.}
\Psi_{5}(q^2) \equiv{}& \hphantom{(m_{d}+m_{u})^2} i\int\,d^4 x e^{iq\cdot x}
  \langle 0\vert T\{\partial^{\mu}A_{\mu}^{(-)}(x),
  \partial^{\nu}A^{(+)}_{\nu}(0)\}\vert 0\rangle ,  \nonumber \\ 
              ={}& (m_{d}+m_{u})^2            i\int\,d^4 x e^{iq\cdot x}
  \langle 0\vert T\{P^{(-)}(x),P^{(+)}(0)\}\vert 0\rangle \ .
\eea
The idea of the standard
analysis \onlinecite{BPR95,becchi81,narison81,dRD} is then to consider
the finite energy sum rules (FESR) generated by integrating products
of the form $t^n\Psi_5(t)$ over the contour shown in
Fig.~\ref{f_contour}.  For $n$ negative the result involves $\Psi_5$
or its derivatives at $t=0$,
% (which is why they are not used to extract 
%the quark masses \onlinecite{BPR95}), 
while for $n$ greater than or
equal to zero, the result is zero.  For sufficiently large radii, $s$,
of the circular portion of the contour, the pseudoscalar two-point
function, and hence also its line integral over the circle, can be
evaluated using perturbative QCD.  Taking the resulting
expressions to the RHS's, one obtains FESR's for the moments of the
spectral function $\rho_5(t)\equiv {\frac{1}{\pi}} {\rm Im}\,
\Psi_5(t)$, on the interval $(0,s)$, for example\onlinecite{BPR95},
\bea 
%% \label {eq:FESR0}
%% \int_{0}^{s} \frac{dt}{t}\, \rho_{5}(t) & = &
%% \frac{N_{c}}{8 \pi^{2}}[m_{u}(s)+m_{d}(s)]^{2}\,
%% s\left\{1+R_{0}(s)\right\} + \Psi_{5}(0);
%% \\ &   & \nonumber \\ 
\label {eq:FESR1}
\int_{0}^{s}dt\, \rho_{5}(t) & = &
\frac{N_{c}}{8 \pi^{2}}[m_{u}(s)+m_{d}(s)]^{2}\, \frac{s^2}{2} 
\left\{ 1+R_{1}(s)+2\frac{C_{4} \langle O_{4} \rangle}
{s^{2}} \right\}; \\
 &   &  \nonumber \\ \label {eq:FESR2}
\int_{0}^{s}dt\, t\, \rho_{5}(t) & = &
\frac{N_{c}}{8 \pi^{2}}[m_{u}(s)+m_{d}(s)]^{2}\,
\frac{s^3}{3} \left\{1+R_{2}(s) - \frac{3}{2}
\frac{C_{6}\langle O_{6} \rangle}{s^3 }\right\},
\eea
where $m(s)$ is the running mass evaluated at the scale $s$, $R_1(s)$
and $R_2(s)$ contain the (higher order in $\alpha_s$) perturbative
corrections, and $C_4\langle O_4\rangle$, $C_6\langle O_6\rangle$
represent the leading non-perturbative corrections, of dimensions four
and six, respectively\onlinecite{SVZ79}.  They are dominated
by the gluon condensate: $C_{4}\langle O_{4}\rangle
\simeq \frac{\pi}{N_c}\langle \alpha_{s} G^2 \rangle$, and the
four-quark condensate which, in the vacuum saturation
approximation, is given by $C_{6}\langle
O_{6}\rangle \simeq (1792/27N_c)\pi^{3}\alpha_{s}\langle \bar{q} q
\rangle^{2}$. Since the contribution of the condensates is negligible, and we 
have no new information to add, 
we simply accept the values quoted by BPR and JM/CPS 
in the remainder of this paper. 

%% TO DO?? OK to leave out the ChPT stuff if we don't quote the LEC
%% combination values, but then we have to remove the G+L reference too.
%% WE should probably also quote the values for the G+L LEC combination
%% that come in $\delta_\pi$ using our model spectral functions.  However
%% this might take some investigation to see how sensitive it is to
%% details of the modeling of the continuum part of the spectral
%% function near $s_th$ (the extra 1/t factor increases the weight for
%% the low t part of the spectral integral).  So do we want to do this or
%% not ???

%% $\Psi_5(0)$ is given, to leading- plus next-to-leading order
%% in the chiral expansion, by\onlinecite{BPR95}
%% \be
%% \Psi_5(0)= 2f_\pi^2m_\pi^2(1-\delta_\pi ) , 
%% \ee
%% where
%% \be
%% \delta_\pi = 4 f_\pi^2 m_\pi^2\frac{1}{F^2}(2L_8^r-H_2^r)
%% \ee
%% with $F$ the $\pi$ decay constant in the chiral limit, and
%% $L_8^r$, $H_2^r$ the renormalized fourth order chiral low-energy
%% constants in the notation of Gasser and Leutwyler\onlinecite{gl85}.

\iffigure
\begin{figure}[t]   %1
\hbox{\hskip15bp\epsfxsize=0.9\hsize \epsfbox {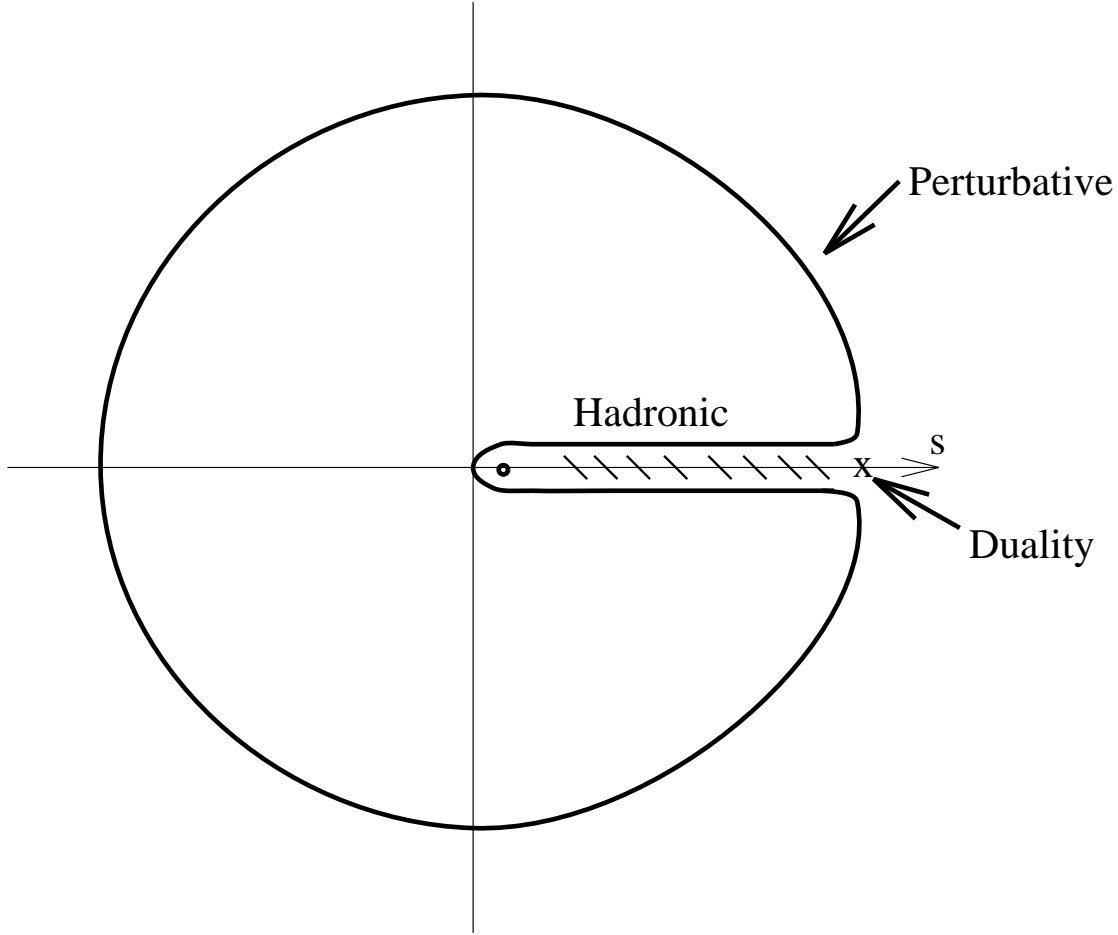}}
\figcaption{The contour integral for the FESR's of the text. The ``hadronic''
integral from $0 \to s$, which includes contributions from the poles
and cuts, is obtained using a model for the continuum portion of the
spectral function, while the integral over the circle at sufficiently
large $s$ ($s>s_0$) is done using the three-loop perturbative result. }
\vskip \baselineskip
\label{f_contour}
\end{figure}
\fi

To extract $m_{u}(s)+m_{d}(s)$, one then needs to input the scale
$s=s_0$, at which one assumes pQCD to have become valid and, second,
experimental and/or model information on the hadronic spectral
function (and hence its moments) below $s_0$.  Having done so, one may
then use either Eq.~(\ref{eq:FESR1}) or (\ref{eq:FESR2}) to extract
$m_{u}(s)+m_{d}(s)$, and from that, the \msbar combination of the
masses at any desired scale $\mu$ using the renormalization group
running. Most sum rule analyses extract
their estimates at $\sqrt{s} \ge 1.7 \GeV$, and then run down to $\mu
= 1 \GeV$. We believe that it is unnecessary to introduce an extra
uncertainty in the estimates by relying on pQCD over this interval
where the running is large. For this reason our final comparisons are
at $\mu = 2 \GeV$.  However, to preserve continuity with existing 
sum rule analyses masses quoted without any argument will always
refer to the \msbar values at $1$ GeV. 

The most up-to-date version of the above analysis was 
performed by Bijnens, Prades and deRafael\onlinecite{BPR95} (BPR),
whose treatment we will follow closely below.  In this analysis, 
BPR have used the three-loop pQCD result 
of Refs.\onlinecite{skl84,skls90} for the pseudoscalar 
two-point function,
employing three active quark flavors with 
$\Lambda^{(3)}_{\overline{MS}} = 300\pm150$ MeV\cite{RPP94},
and the values 
\bea \label{eq:C4O4}
C_{4}\langle O_{4}\rangle & = & (0.08\pm 0.04) {\rm \ GeV}^4 , \\ 
\label{eq:C6O6}
C_{6}\langle O_{6}\rangle & = & (0.04\pm 0.03) {\rm \ GeV}^6 .
\eea
for the non-perturbative, condensate contributions.  For the hadronic
spectral function on the interval $(0,s)$ they include the pion pole,
whose residue is known exactly in terms of $f_\pi$ and $m_\pi$, and a
$3 \pi$ continuum contribution modulated by the $\pi^\prime$ and
$\pi^{\prime\prime}$ resonances. The BPR ansatz is 
\be
\label{eq:ansatz}
\rho_{hadronic}(s) \ = \ \rho_{pole} + F(s) \ \rho^{3\pi}_{\cpt} \Theta(s - 9 m_\pi^2)
\ee
where the ``$3 \pi$ continuum spectral function''
$\rho^{3\pi}_{\cpt}(t)$ is obtained from the leading order, tree-level
chiral perturbation theory (\cpt) result for $\langle
0\vert\partial_\mu A^\mu\vert 3\pi\rangle$, and $F$ is a modulating
factor which accounts for the presence of the $\pi^\prime$ and
$\pi^{\prime\prime}$ resonances. The form of $F$ is taken to be a
superposition of Breit-Wigner terms
\be
F(s) \ = \ A\ { |\sum_i \xi_i / [s-M_i^2 +  i M_i \Gamma_i]|^2 \over 
              { |\sum_i \xi_i / [9m_\pi^2-M_i^2 +  i M_i \Gamma_i]|^2 } } \ , 
\label{eq:Fbpr}
\ee
with $\xi_1 = 1$. 

There remain three unknowns at this point, the overall normalization 
parameter $A$, the relative strength and phase, $\xi_2$, of the two
resonances, and the value of \sdual.  BPR find that, if
they assume $s_0\sim 2$-$3\ \GeV^2$, 
duality can be satisfied for a number of values of $[A,\xi_2]$. Their 
best solution uses the
normalization $A=1$ at threshold, and then fixes $\xi_2$ by demanding
duality between the hadronic ratio 
\be
\label{eq: hadronic ratio}
{\cal R}_{had.}(s)\equiv \frac{3}{2s} \frac{\int_{0}^{s}dt\, t\,
 \rho_{5}(t)} {\int_{0}^{s}dt\,  \rho_{5}(t)},
\ee
and its pQCD counterpart
\be 
\label{eq: QCD ratio}
{\cal R}_{QCD}(s)\equiv \frac{1+R_{2}(s) - \frac{3}{2}
\frac{C_{6}\langle O_{6}\rangle}{s^3}} {1+R_{1}(s)+ 2
\frac{C_{4}\langle O_{4}\rangle}{s^2}}.
\ee
over the interval between the two resonances, $i.e.$ $2.2 \le s \le
3.2 \GeV^2$.  We have reproduced the results of BPR with their choice
of resonance parameters (which differ slightly from those listed in
their published version \cite{BPRprivate}) based on PDB94 \cite{RPP94}, \be
\label{eq:resonanceinput}
M_{1}=1300\ {\rm MeV},\; \Gamma_{1}=325\ {\rm MeV}; \; \; 
M_{2}=1770\ {\rm MeV},\; \Gamma_{2}=310\ {\rm MeV}\ .  
\ee 
Their preferred solution (solution 2) is shown in
Fig.~\ref{f_sfforsd2}.  For $s > s_0 = 3 \GeV^2$ we also plot the
perturbative spectral function (duality constraint) for $\mbar=12$
MeV, their extracted value of the quark mass.  As is evident from the
figure, the rise due to the \piprprime\ is roughly consistent, both in
magnitude and slope, with the perturbative ansatz. This is a
consequence of tuning the normalization and relative phase of the
second resonance, and leads to approximate duality over the range $2.2
{\rm \ GeV}^2<s<3.5 {\rm\ GeV}^2$.  However, the fall-off of the
spectral function on the far side of the \piprprime\ resonance, in
contrast to the rising pQCD solution, shows that,
in order to preserve duality, further
resonances and intermediate states are required to bolster the
BPR ansatz beyond the \piprprime\ peak.

\iffigure
\begin{figure}[t]  %2
\hbox{\hskip15bp\epsfxsize=0.9\hsize \epsfbox {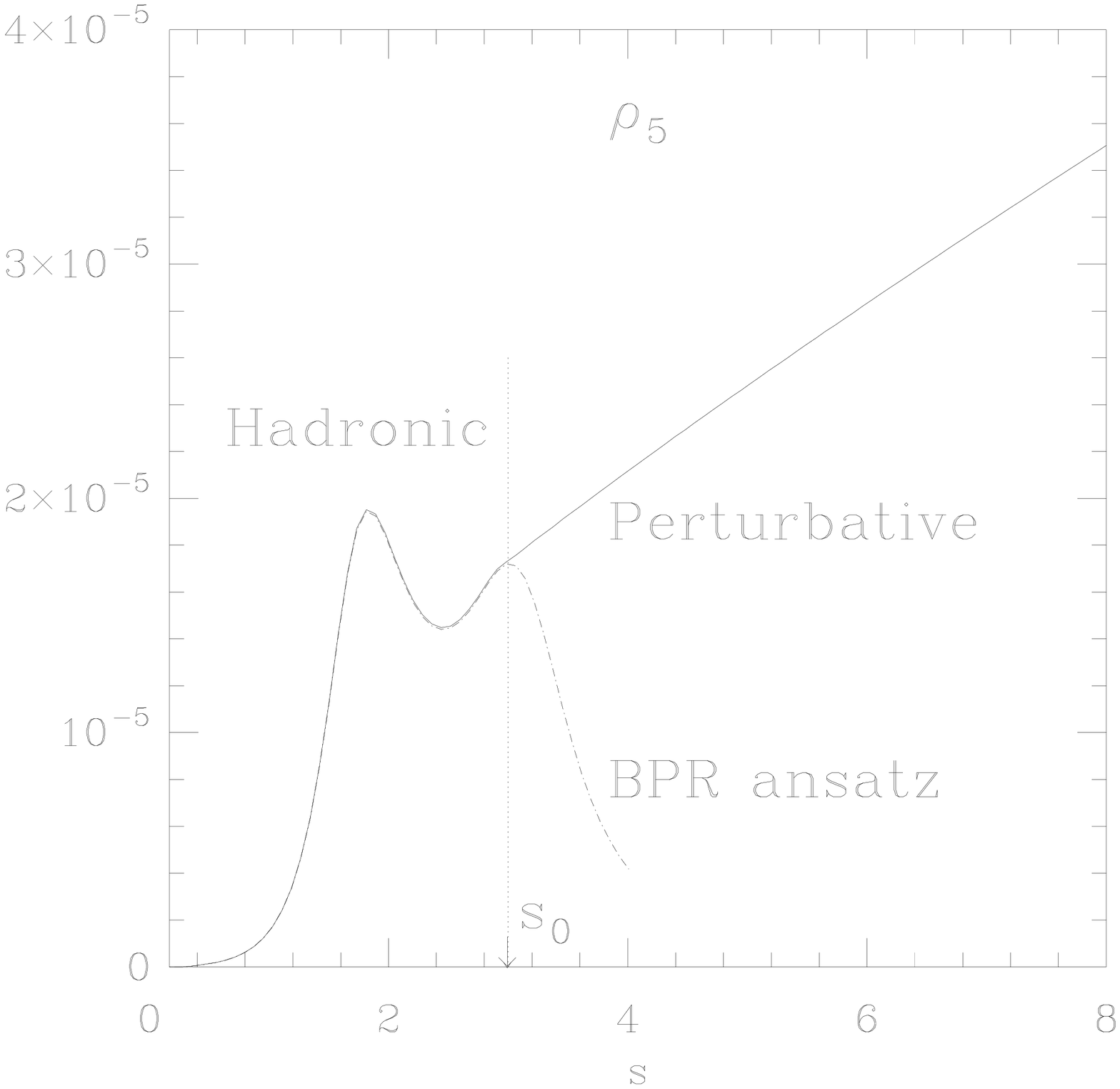}}
\figcaption{The hadronic spectral function assuming that the dual region
begins at the $\pi^{\prime\prime}$ resonance. The location, widths,
and normalization are the same as in the solution found by BPR. The
dashed line is the continuation of the BPR solution for $s > s_0$.
For $s > s_0$ we also show the spectral function required to satisfy
perturbative duality for \summ$=12$ MeV. The two ans\"atze are joined
smoothly by choosing $s_0=3.0$ GeV$^2$. $s_0$ is in GeV$^2$, and
$\rho_5$ in GeV$^4$.}
\vskip \baselineskip
\label{f_sfforsd2}
\end{figure}
\fi

Note that, in the BPR analysis, the 
threshold behavior of the spectral function is not determined
experimentally, but rather obtained from leading order \cpt.  To the
extent that $SU(2)\times SU(2)$ \cpt\ converges well at leading
order, the choice $A=1$ then ensures correct normalization of the
spectral function near $3\pi$ threshold.  However, in the spectral
integral appearing in Eq.~(\ref{eq:FESR1}), which determines the light
quark mass, the contribution of the near-threshold region is
negligible compared to that from the vicinity of the resonance peaks.
Correctly normalizing the spectral function in the resonance region is thus 
much more important than correctly normalizing it 
near threshold.  We will show later, by considering an analogous example
(the correlator of two vector currents),
that the conventional threshold constraint, $A=1$, 
almost certainly leads to a significant {\it overestimate} of the
spectral function in the resonance region.

To summarize, we will investigate the following aspects of the BPR
solution: the uncertainty in the mass extraction produced by
uncertainties in the 3-loop pQCD expression, the reliability of the
overall normalization of the continuum contribution, and the
sensitivity of the results to the value chosen for \sdual, the scale
characterizing the onset of duality with pQCD.  The same issues are also 
relevant to the extraction of $m_s$ using the Ward identity for the 
vector current. Our contention is that
plausible systematic errors in each are such as to lower the
estimates for light quark masses.  
%In particular, we show in Section
%\ref{s:s_vector} that previous methods of normalizing the continuum
%contributions by their value at threshold lead to an overestimate of
%the continuum contribution to the spectral integral, and correcting
%this problem will lead to lower values of the extracted mass.

\section{Convergence of 2-point functions in pQCD}
\label{s:sPQCD}

The pseudoscalar 2-point function is known to three loops in pQCD
\onlinecite{skl84,skls90}. The main issue, in applying this expression
to the problem at hand, is the question of convergence.  If, for
example, we write $1 + R_i$, with $R_i$ as defined in
Eqs.~(\ref{eq:FESR1}) and (\ref{eq:FESR2}), in the form $( 1 + x
\alpha_s/\pi + y (\alpha_s/\pi)^2)$, then the coefficients $x,y$ show
a geometrical growth, $i.e.$ the growth for $R_1$ and $R_2$ is roughly
the same and the average values are $x \approx 6.5$ and $y \approx
46$.  As a result the ${\cal O}(\alpha_s, \alpha_s^2)$ correction
terms are $0.61$ and $0.41$ respectively at $s = 3 \GeV^2$ where
$\alpha_s/\pi \approx 0.1$.  Since these are large, it is important to
estimate the sum of the perturbation series.  One plausible possibility is
to represent the series by the Pad\'{e} $1/(1-0.63)$, in which case the
neglected terms would further increase the pQCD estimate by $\sim
35\%$ and consequently lower the BPR result for $m_u + m_d$ by $\sqrt{
1.35}$, $i.e.$ from $12 \to 10.4 \MeV$. In fact, historically sum-rule
estimates have decreased over time precisely because of the increase
in the pQCD result. Part of the change has been due to the increase in
the value of $\Lambda_{QCD}^{(3)}$ and part due to the large positive
3-loop contribution \cite{BPR95}. If this trend were to continue, and
the unknown higher order terms were to continue to grow geometrically
and contribute with the same sign (as is the case for the scalar channel 
discussed below) then, the extracted
quark mass could be significantly lowered.
%NEED TO DO: Renalyze doing improved perturbation theory, ie do the 
%integral over the circle allowing alpha to vary inside the integral!

The situation in the case of the scalar 2-point function analyzed by
JM/CPS is somewhat better.  A very careful analysis of the stability
of the pQCD expressions and of the choice of the expansion parameter
has been carried out by CPS \cite{CPS96} who include terms up to
$\alpha_s^3$ in the 2-point function and in the running of the
coupling and mass. The pQCD result, after Borel transformation, has the
expansion $( 1 + 4.8 \alpha_s/\pi + 22 (\alpha_s/\pi)^2 + 53
(\alpha_s/\pi)^3)$ \cite{jm95}.  Taking $\alpha_s/\pi \approx 0.1$, as
appropriate for $u = 4 \GeV^2$ with $\Lambda_{QCD}^{(3)}=380 \MeV$, we
find that the difference between the pQCD series and a possible Pad\'{e}
representation, $1/(1-0.48)$, is only about $9\%$. This correction 
would lower the estimate of $m_s$ by $\sim 5\%$, consistent with the 
estimate by JM \cite{jm95}.   

\section{Constraints on \summ from the Positivity of \specfs}
\label{s:spositive}

The fact that the spectral function \specfs is positive definite above
threshold allows us to place rigorous lower bounds on \summ as a
function of \sdual\ \cite{rafael83}.  A weak version of this bound
(labeled ``pole'') is obtained by ignoring all parts of the spectral
function except for the pion pole, whose contribution to the integral
in Eq.~(\ref{eq:FESR1}) is $2f_\pi^2m_\pi^4$ (Eq.~(\ref{eq:FESR2})
produces a much less stringent bound and hence is not considered
further).  One then finds, {\it assuming the validity of the input
3-loop pQCD result},
\be
\label{eq:lower0}
(m_u(s) + m_d(s))^2 \ \ge \ {2 f_\pi^2 m_\pi^4 \over { 
\frac{N_{c}}{8 \pi^{2}}\, \frac{s^2}{2}
\left\{ 1+R_{1}(s)+2\frac{C_{4} \langle O_{4} \rangle} {s^{2}} \right\} } }
\ee
where $s$ is the upper limit of integration in Eq.~(\ref{eq:FESR1}).
A stronger constraint (labeled ``ratio'') is obtained by noting that,
for \specf$\geq 0$,
\be
\label{eq:lowerratio}
{ {\int_{s_{th}}^{s}dt\, t\, \rho_{5}(t) } \over 
  {\int_{s_{th}}^{s}dt\,     \rho_{5}(t) }       } \ \le \ s
\ee
where $s_{th}$ denotes the $3\pi$ threshold value.  The bound is saturated
when the entire spectral strength is concentrated as a delta function at $s$. 
If $s$ in
Eq.~(\ref{eq:lowerratio}) is assumed to be in the dual region, this
turns out to place considerably stronger constraints on
$m_{u}(s)+m_{d}(s)$.  To see this, note that the LHS of the inequality in 
Eq.~(\ref{eq:lowerratio}) is, using Eqs.~(\ref{eq:FESR1}) and (\ref{eq:FESR2}),
\be
\label{eq:ratiodetail}
\frac{\frac{N_{c}}{8 \pi^{2}}[m_{u}(s)+m_{d}(s)]^{2}\,
\frac{s^3}{3} \left\{1+R_{2}(s) - \frac{3}{2}
\frac{C_{6}\langle O_{6} \rangle}{s^3 } \right\} - 2 f_\pi^2 m_\pi^6 }
{\frac{N_{c}}{8 \pi^{2}}[m_{u}(s)+m_{d}(s)]^{2}\, \frac{s^2}{2} 
\left\{ 1+R_{1}(s)+2\frac{C_{4} \langle O_{4} \rangle}
{s^{2}} \right\}- 2f_\pi^2 m_\pi^4}\ .
\ee
From the expression~(\ref{eq:ratiodetail}) we see that, if we start
with a large value of $m_u(s)+m_d(s)$ and begin to lower it, keeping
$s$ fixed, the inequality (\ref{eq:lowerratio}) will be violated
before we reach the value of $m_u(s)+m_d(s)$ corresponding to the pion
pole saturation of the spectral function at which point the denominator 
in (\ref{eq:ratiodetail}) vanishes).  Thus the inequality
(\ref{eq:lowerratio}) provides a more stringent (larger) lower bound
on the extracted quark mass.  This is illustrated in
Fig.~\ref{f_lowerbounds} where the dependence of $(m_{u}+m_{d})_{min}$
on $s$ for both of the above constraints is shown.

The ``ratio'' curve shows that if one assumes $s_0\simeq 2.5$ GeV$^2$,
as in the BPR analysis, 
then \summ$\geq 10$ MeV.  The fact that the BPR result for
the mass extraction, \summ$\simeq 12$ MeV, is close to this lower
bound is a reflection of the fact that the spectral strength
is concentrated in the region close to the assumed onset of duality.
Such a feature is, in fact, rather natural since \sdual\ is chosen to
coincide with the \piprprime\ peak. However, if \sdual is considerably
larger than $3$ GeV$^2$ (to alleviate the problem of large ${\cal O}(
\alpha_s ,\ \alpha_s^2)$ corrections to $R_i$ at $s\sim 3\ \GeV^2$
discussed above) then considerably smaller masses are allowed
by the ``ratio'' constraint, as is evident from the figure.
Furthermore, one would, in fact, expect masses not much greater than
the ``ratio'' bound to be favored in all cases where spectral
functions are characterized by resonance modulation of a rising
continuum phase space background and have their spectral strength
concentrated in the region near $s_0$.  

\iffigure
\begin{figure}[t]    %3
\hbox{\hskip15bp\epsfxsize=0.9\hsize \epsfbox {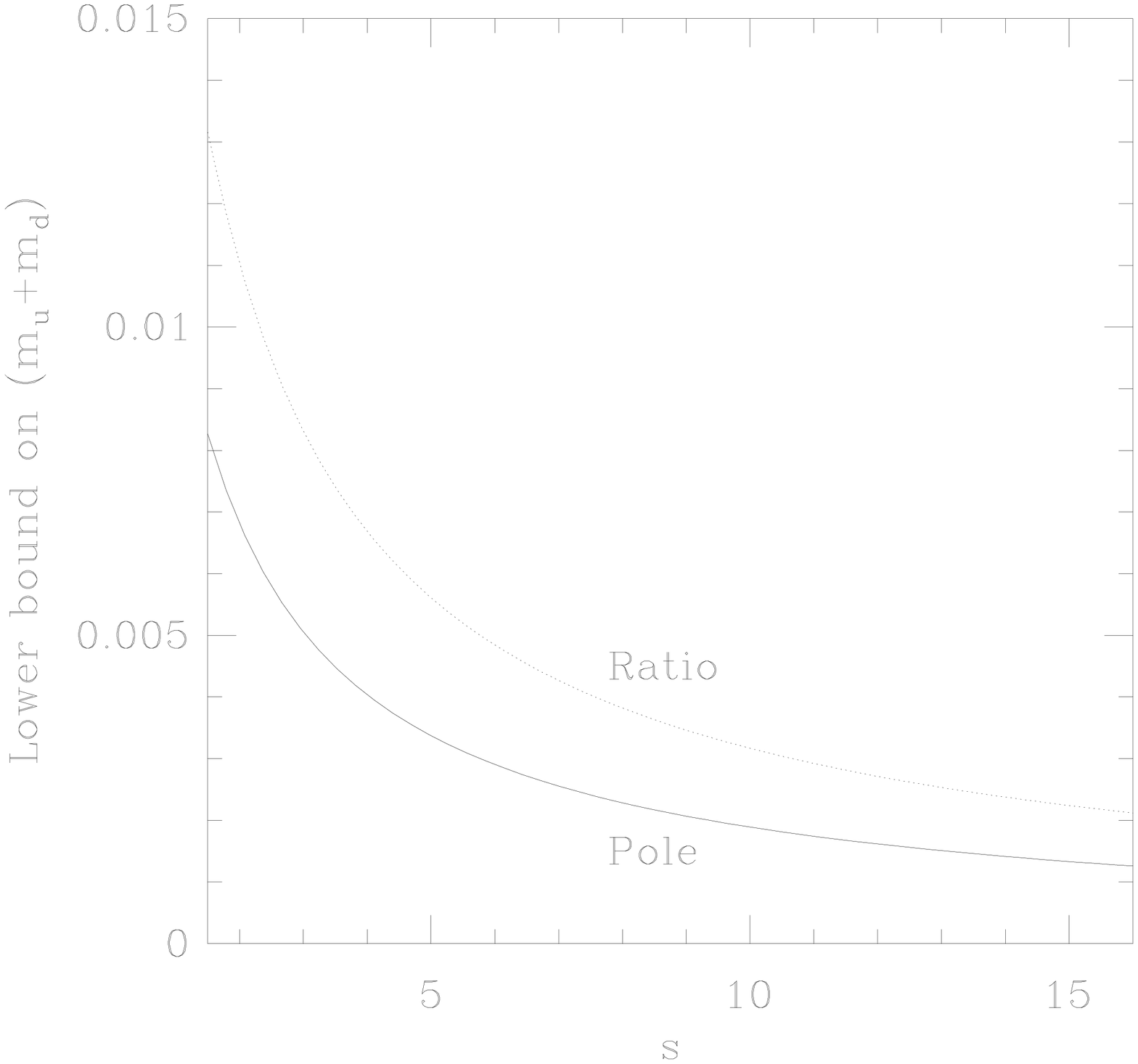}}
\figcaption{The lower bounds on $m_u + m_d$ as a function of $s$. These are 
obtained by saturating the spectral function with the 
pion pole contribution, and from the ``ratio''
method described in the text.  $s$ is in GeV$^2$ and $m_u+m_d$ in GeV.}
\vskip \baselineskip
\label{f_lowerbounds}
\end{figure}
\fi

These bounds make it clear that the value of the quark mass extracted
from FESR will tend to be very strongly correlated with assumptions
about the appropriate value of \sdual.  In addition, it will, of
course, depend on the details of the hadronic spectral function from
$3\pi$ threshold up to $s_0$, which are, at present, not
experimentally determined.  Since the perturbative \specfpunc\ is
known up to the overall normalization, which is given by the quark
mass, one test of the validity of the phenomenological ansatz for the
hadronic spectral function would be to show that the results for the
quark mass remained stable under variations of the upper limit of
integration, $s$, in Eqs.~(\ref{eq:FESR1}) and (\ref{eq:FESR2}).  This
test, however, is meaningful only if one already knows that the values
of $s$ being employed are greater than $s_0$. Unfortunately, the lack
of experimental information on the hadronic $\rho_5(s)$ precludes the
possibility of making such a test. In the next Section we construct
plausible spectral functions, all satisfying duality, corresponding to
a range of possible values for $s_0$ lying between $3$ and $10\
\GeV^2$, by including higher resonances in the $3 \pi$ channel.  These
models illustrate how, in the absence of experimental information, the
uncertainty in \summ\ might be as large as a factor of $2$ if
considerably higher values of $s_0$ are chosen.

\section{Plausible Spectral Functions \specfs in the Dual Region}
\label{s:sguess}

The assumption of duality places constraints on the form of the
spectral function, $\rho_5(s)$.  Below $s=s_0$, these constraints
amount only to the determination of certain moments of the spectral
function on the interval $(s_{th},s_0)$, and hence are not
particularly strong.  In fact, as we illustrate below, the constraints of
Eqs.~(\ref{eq:FESR1}) and (\ref{eq:FESR2}) allow considerable freedom
in the choice of $\rho_5(s)$ for $s < s_0$. For $s>s_0$, in contrast,
duality determines the ``average'' $\rho_5(s)$, $i.e.$ averaged over
some suitable region of $s$. This average value is given by the
perturbative \specfs, which can be obtained straightforwardly by
differentiating the RHS of either Eq.~(\ref{eq:FESR1}) or
(\ref{eq:FESR2}).  We evaluate these derivatives numerically using
either of the two forms, which of course give consistent results.
Even if one eliminates the running masses by matching the ratios of
Eqs.~(\ref{eq: hadronic ratio}) and (\ref{eq: QCD ratio}) for $s>s_0$,
it is easy to show that the resulting equation completely determines
the perturbative \specfspunc, up to an overall multiplicative factor,
for all $s>s_0$.  The result of the duality constraints, in either
form, is that \specfs must be a monotonically increasing function of
$s$, for $s$ in the duality region.  Numerically we find that this
function is approximately linear as illustrated in Fig.~\ref{f_sfforsd2}.

The hadronic spectral function in this channel is not known
experimentally.  It receives contributions not only from the pion and
its resonances, but also from the resonant and non-resonant portions
of the $3\pi, 5\pi, 7\pi, K\bar K \pi, \ldots$, 
$N \bar N$, $\ldots$ intermediate
states.  Experimentally, only the \piprime\ and \piprprime\ have been
observed as distinct resonances\onlinecite{RPP96}. Even so, their
decay constants are not known experimentally, and hence the
normalization of their contributions to the spectral function 
have to be treated as free parameters. The number of
multi-particle intermediate states one has to consider, moreover,
grows with $s_0$, as does the problem of separating their resonant and
non-resonant portions.  From dimensional arguments, the
contribution of these various intermediate states will grow linearly
at sufficiently large $s$.  In the region of resonances, the
resonances will modulate the cut contribution, and the hadronic
spectral function is expected to match the pQCD behavior only after an
average over some interval of $s$.  This averaging is crucial if the
resonances are narrow and isolated. Alternately, if the widths of
subsequent resonances, for example $\pi^{\prime\prime\prime}$ and
$\pi^{\prime\prime\prime\prime}$, become much greater than the
resonance separation, then the overlap of resonances can provide the
monotonically rising behavior required by duality, and averaging is
not crucial.

We illustrate these points by constructing ans\"atze for the hadronic
spectral function which are of the form used by BPR, $i.e.$, involving
resonance modulation of the continuum $3\pi$ background.  To explore
values of $s_0$ as large as $10$ GeV$^2$ with the ansatz above, we
include pseudoscalar resonances with masses as large as $\sim
\sqrt{10}$ GeV.  For the first two such resonances, the $\pi^\prime
(1300)$ and $\pi^{\prime\prime}(1800)$, 
we use the 1996 Particle Data Group values for the masses and widths.
For the remaining two resonances, the $\pi^{\prime\prime\prime}$ and
$\pi^{\prime\prime\prime\prime}$, expected in this range, we are
guided by model predictions.  The $\pi^{\prime\prime\prime}$ resonance
is typically expected to lie around $2400$ MeV in models constrained
by the lower part of the meson spectrum\onlinecite{godfrey}. In
addition, the ${}{^3P_0}$ model\onlinecite{3p0new}, which has proven
reasonably successful in estimating decay widths\onlinecite{3p0},
predicts a width for the $\pi^{\prime\prime\prime}(2400)$ between
$700-1900$ MeV\onlinecite{godfrey,godfreypc} depending on how the
relativistic effects are treated.  The approach leading to $700$ MeV
gives $300$ MeV for the width of the \piprprime\ which is larger than
the experimental value of $212(37)$ MeV.  We, therefore, assume the
lower limit $700$ MeV for the width in this study, even though this
may be an overestimate.  Similarly, we assume that
$\pi^{\prime\prime\prime\prime}$ lies at $3150$ MeV with a width of
$900$ MeV.  In short, we choose 
\bea
\label{eq:R34input}
M_{1} &=& 1300\ {\rm MeV},\; \Gamma_{1}=325\ {\rm MeV}; \; \; 
M_{2}  =  1800\ {\rm MeV},\; \Gamma_{2}=212\ {\rm MeV}; \nonumber \\
M_{3} &=& 2400\ {\rm MeV},\; \Gamma_{3}=700\ {\rm MeV}; \; \; 
M_{4}  =  3150\ {\rm MeV},\; \Gamma_{4}=900\ {\rm MeV}.
\eea
The decay constants of all of these resonances are unknown, and will
therefore be treated as free parameters.  The limitations of such a
truncated spectral function are obvious; however, it should be noted
that, because we have allowed ourselves some phenomenological freedom
in treating the strengths and widths of the last two resonances, our
ans\"atze for the spectral function can also be thought of as providing
an approximate means of representing a combination of resonant and
non-resonant effects.  Our aim is, in any case, to simply demonstrate
how the piling up of resonances can give the $pQCD$ behavior, and the
nature of plausible spectral functions for which the ``extracted''
quark mass is, as for the BPR case, rather close to the value given by
the ``ratio'' bound.

For the resonance modulated spectral function we adopt, following BPR, 
the ansatz
\be
\rho_5(s)=F(s)\ \rho^{3\pi}_{\cpt}(s)
\label{eq:modulation}
\ee
where $\rho^{3\pi}_{\cpt}(s)$ is the spectral function corresponding
to the leading order, tree-level \cpt\ result for {\me}, and
\be
F(s) \ = \ A  { \sum_i c_i M_i \Gamma_i / [(s-M_i^2)^2 +  M_i^2 \Gamma_i^2] \over 
              { \sum_i c_i M_i \Gamma_i/ [(s_{th}-M_i^2)^2 + M_i^2 \Gamma_i^2] } } \ . 
\label{eq:Fkimraj}
\ee
The sum in Eq.~(\ref{eq:Fkimraj}) runs over the appropriate number of
resonances, depending on \sdual as described below, with relative
strengths $c_i$. The parameter $A$ is the overall normalization of the
resonance contribution to the continuum part of the spectral function
at $3\pi$ threshold.  We have taken the $c_i$ to be real, in order to
simplify the task of searching for suitable spectral functions,
whereas BPR, who use a slightly different form for F, as given in
Eq.~(\ref{eq:Fbpr}), with just the first two resonances, allow the
relative strength of the two resonances to be complex.

We display a series of spectral functions in Fig.~\ref{f_sfcomp}, all
satisfying duality and constructed by employing up to four resonances
in the ansatz above.  The values for $A$, $\{ c_i\}$, $s_0$ and
$\summ$ used in the construction are given in Table~\ref{t_fitcomp}.
As one can see from the figure, there exist perfectly plausible
spectral functions corresponding to \summ$=12, 9, 8,$ and $6$ MeV. The
first case ($s_0 = 3 \GeV^2$, $\summ=12 \MeV$) is the BPR solution
discussed before.  The second case ($s_0 = 5.7 \GeV^2$, $\summ=9
\MeV$) corresponds to including three resonances and matching to the
duality solution at the top of the third resonance.  The assumption
here is that the third and higher resonances merge to produce the dual
solution above this point.  The matching in the third case ($s_0 = 8
\GeV^2$, $\summ=8 \MeV$) is at the beginning of the rise of the fourth
resonance, while in the fourth case ($s_0 = 10 \GeV^2$, $\summ=6
\MeV$) we match at the top of the fourth resonance.  In cases where we
match at the peak of a resonance, the dual region actually appears to
begin somewhat below the input value of \sdual .  This is because the
slope of the rising side of the last resonance tends to match
reasonably well the slope of the pQCD version of $\rho_5(s)$.  

\iftable
\begin{table} %1
\caption{The parameters used to generate the spectral functions shown in 
Fig.~\ref{f_sfcomp}. The normalization at threshold 
$A$ and the relative weights $c_i$ assigned 
to the resonances are defined in Eq.~(\ref{eq:Fkimraj}). \looseness=-1}
\vskip 6pt
\input {t_fitcomp}
\vskip -12pt plus 10pt
\label{t_fitcomp}
\end{table}
%% 7/23/96
\fi

\iffigure
\begin{figure}[t]   %4
\hbox{\hskip15bp\epsfxsize=0.9\hsize \epsfbox {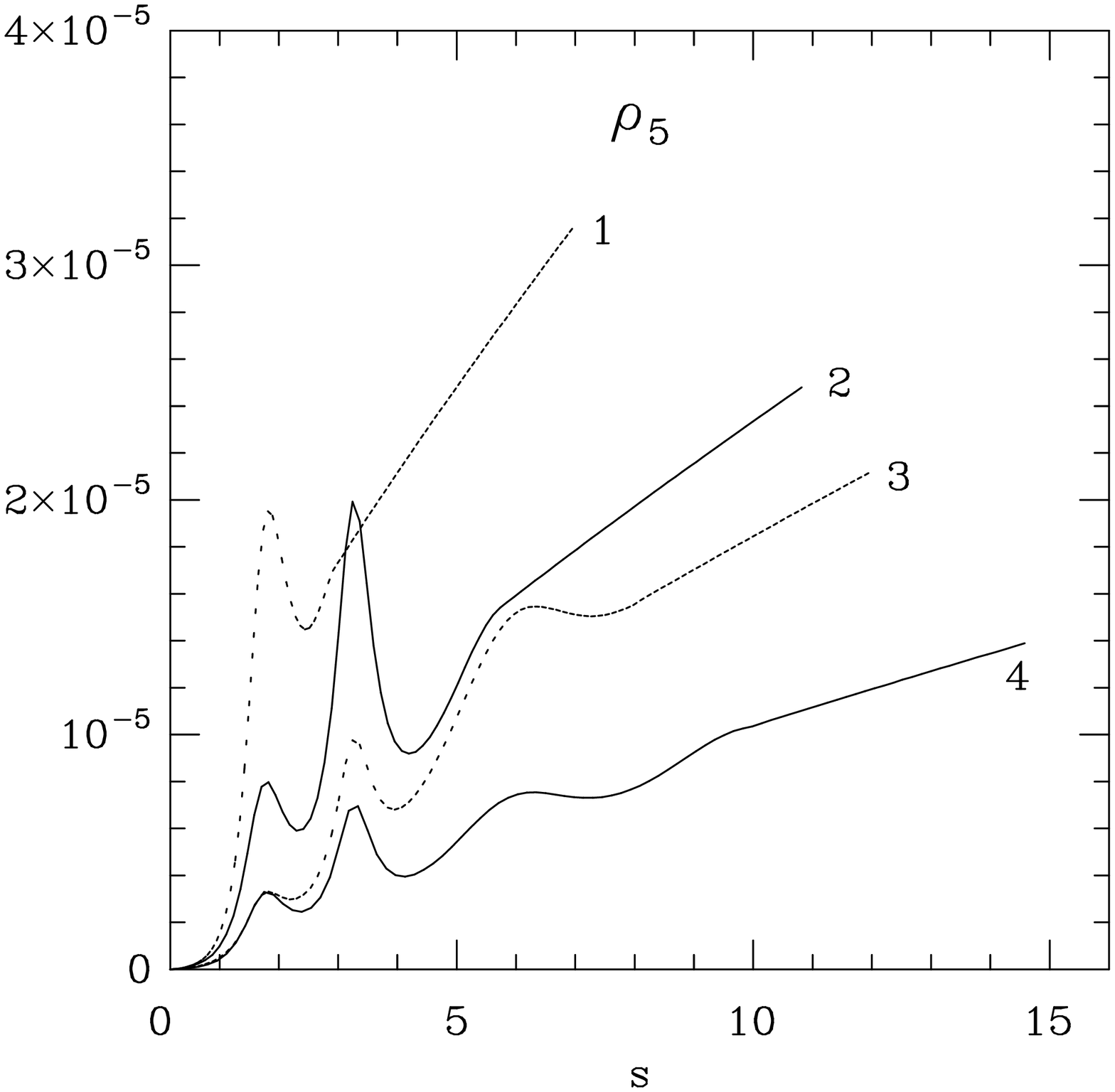}}
\figcaption{Four examples of the hadronic spectral function, assuming 
different resonance structure and point of matching to the
perturbative solution.  Units are as in Fig.~\ref{f_sfforsd2}. 
The locations and widths of the resonances used
are given in the text.  The normalization $A$ and the relative
strengths $c_i$ for the four cases are given in Table~\ref{t_fitcomp},
along with the values of $s_0$ and $\summ$ used to derive the
perturbative solution.}
\vskip \baselineskip
\label{f_sfcomp}
\end{figure}
\fi

These spectral functions are, by construction, perfectly dual for
$s>s_0$.  Duality also requires the low-energy ($s<s_0$) part of
\specfs to have the correct moments to satisfy Eqs.~(\ref{eq:FESR1}),
(\ref{eq:FESR2}).  However, the constraint of duality does not lead to a
unique solution. Experimental data (decay constants) are needed to fix
the overall normalization $A$ and the relative weights $c_i$.  We
illustrate this point in Fig.~\ref{f_sfforsd10} by constructing three
spectral functions that differ for $s < s_0$. In all three cases $s_0$
and the input $\summ$ in the pQCD expression are fixed to be the same,
while the values of parameters $A$ and $c_i$ are as defined in
Table~\ref{t_fitssd10}.  The corresponding output values for $\summ$
are shown in Fig.~\ref{f_mqfor10}.  As expected they converge to the
input value in the dual region.

There are three features of these ans\"atze that should
be noted.  First, the
value of \mbar\ decreases with $s_0$ in a manner very similar to the
``ratio'' bound. This is because in each case the spectral function is
stacked up towards $s_0$. Second, we find that, to produce spectral
functions corresponding to values of {\summ} only a few MeV above the
``ratio'' bound, the threshold normalization parameter, $A$, has to be
decreased with increasing $s_0$. In the next section we will show that
values of $A$ significantly smaller than $1$ are, in fact, to be
expected, based on a consideration of the analogous vector current
correlator, for which the normalization in the resonance region is
known experimentally.  Third, the $c_i$ are large. It is not clear, 
{\it a priori}, if
this should be considered unreasonable or not. For example, in the
narrow width approximation the $c_i$ would scale as 
$ \sim (f_i^2 M_i^4)/(f_\pi^2 M_\pi^4)$
and thus have an explicit dependence on $M_i^4$.  (The BPR model
spectral functions, being even larger than ours, of course, correspond
to even larger \piprime\ and \piprprime\ decay constants.)
Moreover, by leaving the
normalizations as free parameters we are potentially incorporating
other non-resonant background effects.  Ultimately, this issue can only
be resolved by appeal to experimental data which, unfortunately, is
not available at present.

%% NOTES: For hydrogen like wave functions the prob at the origin 
%% falls as $1/n^3$ for the $n$ radial state. However, we these states 
%% are not described by the NR limit.

\iftable
\begin{table} %2
\caption{The parameters used generate the plots shown in 
Fig.~\ref{f_sfforsd10}.  The values of $s_0$ and $\summ$ have been
fixed to $s_0 = 10\GeV^2$ and $\summ=6 \MeV$ respectively in each of
the three cases. \looseness=-1}
\vskip 6pt
\input {t_fitssd10}
\vskip -12pt plus 10pt
\label{t_fitssd10}
\end{table}
\fi
%% 7/23/96

%% The convergence behavior of the pQCD solution is not very helpful since
%% in the expansion $A( 1 + \lambda x + \lambda^2 y)$ the constants are
%% $x = 6.7$ and $y = 49.8$ and $\lambda = \alpha_s/\pi $ changes from
%% $0.091 \to 0.07$ between $3 $ and $ 10 \GeV^2$. What
%% is clear is that using pQCD at $s$ as low as $\sim 2.5 \GeV^2$ may lead to 
%% a large uncertainty in the determination of quark masses.

%The criteria we would like to use is that given a choice of $s_0$, the
%hadronic spectral function should agree (in the sense of an average)
%with the $pQCD$ solution for all $s > s_0$. As an illustration of the
%difficulty in making such a criteria practical, we compare the $pQCD$
%result for the BPR and our (case 4 in Fig.~\ref{f_sfcomp}) ansatz in
%Fig.~\ref{f_3picomp}. One finds, for example, for the BPR solution,
%$pQCD$ gives $\rho_5(s=10 \GeV^2) = 4.2 \times 10^{-5}$, whereas the
%$3 \pi$ contribution $\rho_{\cpt}(s)$ is only $5.67 \times 10^{-6}$.
%Thus, the resonance modulation factor $F(s)$ and the multiparticle
%states neglected in Eq.~(\ref{eq:modulation}) have to account for
%$\approx 85\%$ of the answer. On the other hand, the enhancement is a
%much smaller for our solution with only resonance modulation, $i.e.$
%the factor is $\sim 1.85$ at $s=10 \GeV^2$.  The true answer probably
%lies somewhere in between.

The bottom line of the above discussion is that since both the correct
value for the location of the onset of duality with pQCD,
and the correct form of the hadronic spectral function, are at present
unknown, the value of \summ\ extracted using FESR can easily vary by a
factor of $2$. As we have pointed out, using $s_0
\sim 3$ GeV$^2$ leads to a perturbation series in which the $\alpha_s$
and $\alpha_s^2$ terms are large.  As soon as one allows
significantly larger values of $s_0$, in order to alleviate this
problem, however, considerably smaller values of the extracted quark mass are
possible.  We will now, furthermore, argue that the conventional method
of normalizing the continuum part of the spectral function tends
to produce significant overestimates of the resonance contributions,
and hence also significant overestimates of the extracted quark masses.
%We will explain why this is the case based
%on an understanding of how resonance effects manifest themselves in
%the chiral expansions of low-energy observables, as calculated in
%{\cpt}, and will then use this understanding to provide an alternate
%procedure for normalizing resonance contributions for those cases
%where the $O(p^4)$ \cpt\ corrections to the threshold behavior are
%known.

\iffigure
\begin{figure}[t]  %5
\hbox{\hskip15bp\epsfxsize=0.9\hsize \epsfbox {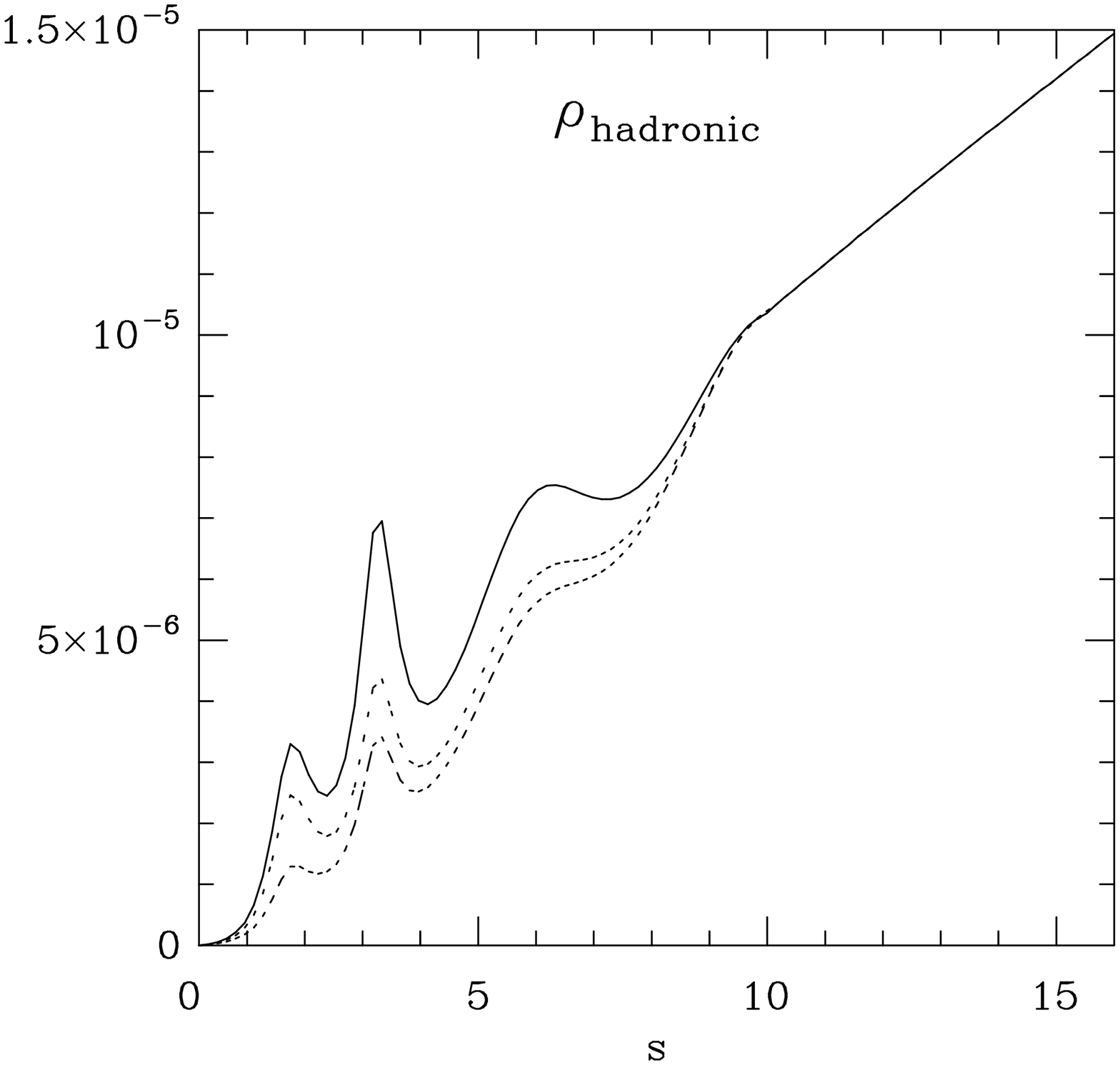}}
\figcaption{The hadronic spectral function assuming four resonances with 
the quantum numbers of the pion.  Units are as in Fig.~\ref{f_sfforsd2}. 
The locations and widths of the
resonances are given in the text. The normalization $A$ and relative
strengths $c_i$ for the three cases are given in
Table~\ref{t_fitssd10}.  These have been adjusted to make the hadronic
form join smoothly to the duality ansatz at $s_0 = 10 \GeV^2$.  The
solid line corresponds to $Case\ 1$, the dotted line to $Case\ 2$, and
the dashed line to $Case\ 3$. }
\vskip \baselineskip
\label{f_sfforsd10}
\end{figure}
\fi

\iffigure
\begin{figure}[t]   %6
\hbox{\hskip15bp\epsfxsize=0.9\hsize \epsfbox {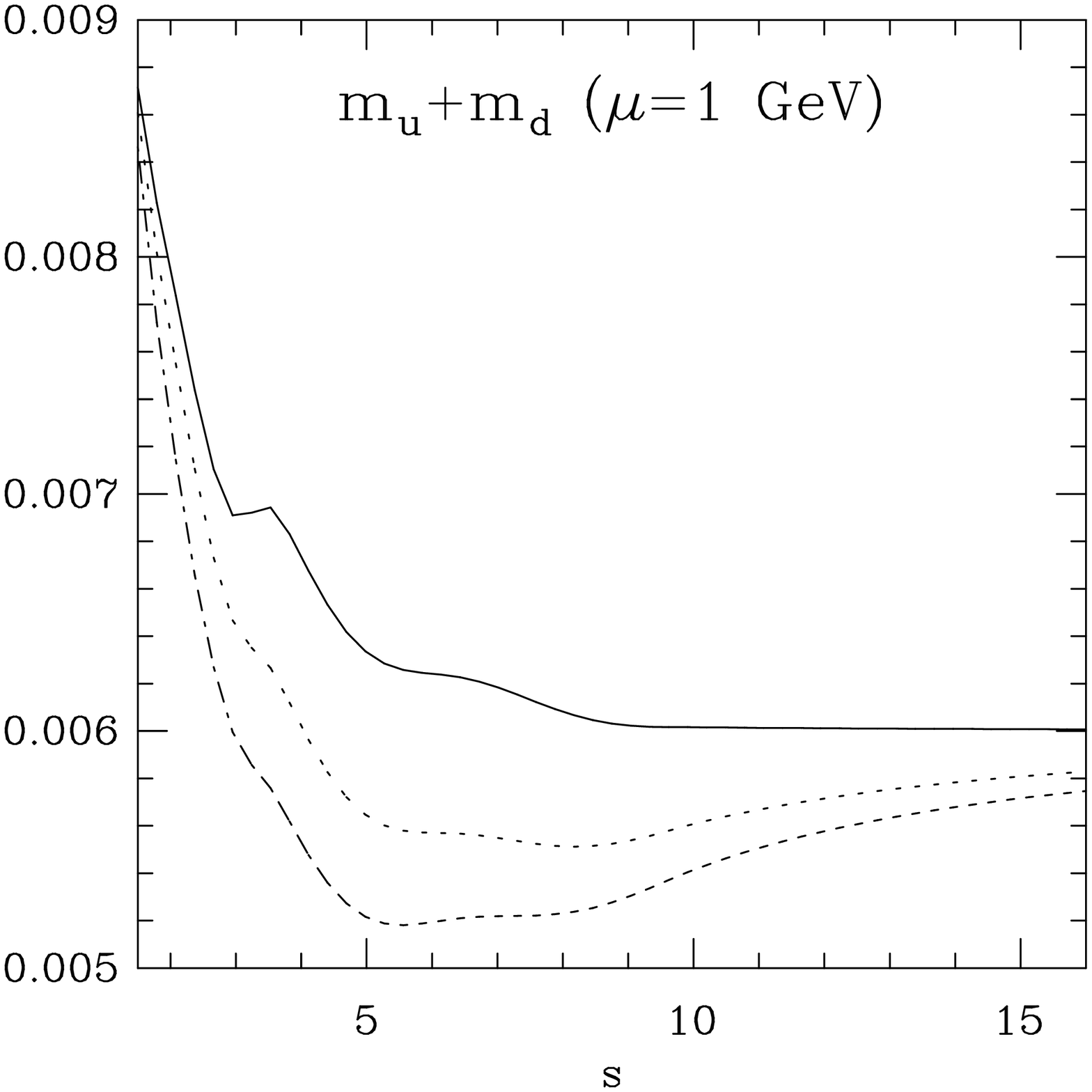}}
\figcaption{The output values of $m_u + m_d$ (run to $\mu=1\GeV$) for the 
three cases of the spectral function shown in Fig.~\ref{f_sfforsd10}.
$s$ is in GeV$^2$ and $m_u+m_d$ in GeV. }
\vskip \baselineskip
\label{f_mqfor10}
\end{figure}
\fi

%% \iffigure
%% \begin{figure}[t]   %7
%% \hbox{\hskip15bp\epsfxsize=0.9\hsize \epsfbox {f_3picomp.ps}}
%% \figcaption{A comparison of the BPR solution and our ansatz (case 4 in
%% Fig.~\ref{f_sfcomp}) with $\rho_{\cpt}^{(3\pi)}$. The BPR solution is
%% supplemented by the duality result for $s \ge 3 \GeV^2$.  Units are
%% as in Fig.~\ref{f_sfforsd2}.}
%% \vskip \baselineskip
%% \label{f_3picomp}
%% \end{figure}
%% \fi

\section{Normalization of the resonance-modulated spectral function}
\label{s:s_vector}

The assumption that a given spectral function may be written as a
Breit-Wigner resonance modulation of a continuum phase space factor,
as exemplified in Eqs.~(\ref{eq:ansatz}) and (\ref{eq:Fbpr}), is valid in
the vicinity of a narrow resonance. We will assume that this ansatz,
used by both BPR and JM/CPS (see section \ref{s:s_JM}) in the
pseudoscalar and scalar channels respectively, is a good
approximation.  This fixes the general form of the spectral function
but, of course, does not fix the magnitude in the resonance region,
since the relevant pseudoscalar and scalar decay constants (with the
exception of $f_\pi$) are not known experimentally.  Both BPR and
JM/CPS deal with this problem by assuming that resonance dominance of
the relevant spectral function continues to hold all the way down to
continuum threshold.  Thus, for example, the overall scale of the BPR 
and JM/CPS 
ans\"atze for the continuum part of the spectral function 
is obtained by
choosing $A=1$, {\it i.e.}, by assuming that the tails of the
resonances reproduce the {\it full} threshold spectral function. BPR, 
in the absence of experimental data, use the tree-level {\cpt}
expression for the spectral function in the 
threshold region.  The JM/CPS
treatment differs only in that they normalize the sum-of-resonances
ansatz at $K\pi$ threshold using experimental data (the scalar form
factor at threshold is computed using the Omnes representation 
with experimental $K\pi$ phase shifts as input.)

The second key point in the JM/CPS ansatz for spectral function is the
assumption that one can take the ``standard'' s-wave $s$-dependent
widths for the resonance contributions. This assumes that the
effective coupling of the strange scalar resonances to $K\pi$ is
momentum-independent over the whole kinematic range relevant to the
spectral integral.  We will now show that the combination of this
assumption and of the resonance-saturation-at-threshold can fail badly
by studying its exact analogue in the isovector vector channel.  In
fact, in the vector channel, the analogous set of assumptions produces
a significant overestimate of the spectral strength in the region of
the resonance peak.

Consider, therefore, the vector correlator
\bea
\label{eq:6vec}
\Pi^{\mu\nu}_{33}(q^2)&\equiv & \left(q_\mu q_\nu -q^2g_{\mu\nu}\right)
\Pi_{33}(q^2)\nonumber \\
&=& {\rm i}\, \int d^4x e^{{\rm i}q.x} \langle 0\vert T\{V^\mu_3(x)
V^\nu_3(0) \} \vert 0\rangle\ ,
\eea
where $V^\mu_3$ is the $I=1$ vector current.  In the narrow width
approximation, the $\rho$ contributions to the spectral function of
$\Pi_{33}$, $\rho_{33}$, at the $\rho$ peak, is known in terms of the
$\rho$ decay constant, $F_\rho = 154{\rm \ MeV}$,
\be
\label{eq:6rho}
\left[ \rho_{33}(m^2_\rho )\right]_\rho = {\frac{F^2_\rho}{\pi \Gamma_\rho
m_\rho}}=0.0654\ .
\ee
Let us now apply the analogue 
of the BPR and JM/CPS ans\"atze to the vector channel,
by assuming the (trial) spectral function to be given by
\be
\label{eq:6trial}
\rho_{33}^{trial}(s)=\left[ {\frac{1}{48\pi^2}}\left( 1-{\frac{4m^2_\pi}
{s}}\right)^{3/2} \theta (s-4m^2_\pi )\right]\left[ {\frac{c_\rho^{BW}(s)}
{c_\rho^{BW}(4m^2_\pi )}}\right]
\ee
where the quantity in the first set of square brackets is the leading
order \cpt\ expression for the spectral function\cite{gk95}, and
\be
\label{eq:6bw}
c_\rho^{BW}(s) = \frac{sm_\rho\Gamma_\rho /\left[ m^2_\rho -4 m^2_\pi \right]
^{3/2}}{\left[ (m^2_\rho -s)^2+\Gamma_\rho (s)^2m_\rho^2\right]}
\ee
which follows from employing the p-wave s-dependent width
\be
\label{eq:6edw}
\Gamma_\rho (s)=
{\frac{m_\rho \Gamma_\rho }{\left( m_\rho^2-4m_\pi^2 \right)^{3/2}}}
\left[ 1-{\frac{4 m_\pi^2}{s}}\right]^{3/2}s \ .
\ee
We have chosen this form of the width in analogy to the ``standard'' s-wave
$s$-dependent width of JM/CPS. This ansatz assumes that
the effective coupling of the $\rho$ to $\pi\pi$ has the minimal
form $g_{\rho\pi\pi}\rho_\mu \left( \pi^+\partial^\mu \pi^-
-\pi^-\partial \pi^+\right)$ with $g_{\rho\pi\pi}$ independent
of momentum over the relevant kinematic range.
The threshold factor $\left[ 1-{4 m_\pi^2}/{s} \right]^{3/2}$ 
in the numerator of Eq.~(\ref{eq:6edw})
has been separated out explicitly in writing
Eq.~(\ref{eq:6trial}).  The ansatz Eq.~(\ref{eq:6trial}) then implies 
\be
\label{eq:6peak}
\rho_{33}^{trial}(m_\rho^2)={\frac{1}{192\pi^2}}\frac{\left( m_\rho^2-
4m^2_\pi \right)^{7/2}}{m_\rho^3\Gamma_\rho^2 m_\pi^2}= 0.27\ ,
\ee
a factor of $4.1$ too large.  Had we instead used the normalization
given by the full next-to-leading order \cpt\ expression \cite{gk95}, 
(which matches well to experimental data near threshold),
\be 
\label{eq:6chpt}
\rho_{33}^{\cpt}(s)= {\frac{1}{48\pi^2}}\left( 1-{\frac{4m^2_\pi}
{s}}\right)^{3/2} \theta (s-4m^2_\pi )\left[ 1+{\frac{4L_9^r(\mu )s}{f_\pi^2}}
+\cdots\right] \ ,
\ee
the peak height would be further increased by a factor of $1.28$, the
correction being dominated,
for $\mu\sim m_\rho$, by the term in the square brackets in
Eq.~(\ref{eq:6chpt})
involving the ${\cal O}(q^4)$ renormalized low-energy constant
(LEC), $L_9^r$,
\be
\label{eq:6norm}
{\frac{4L_9^r(m_\rho )(4m_\pi^2)}{f_\pi^2}}=0.24\ , 
\ee
where we have used $L_9^r(m_\rho) = 0.0069(2)$ \cite{BKM95}, and
$+\cdots$ refers to loop contributions whose form is not important in
what follows.  Note that, since it is the next-to-leading order
expression, Eq.~(\ref{eq:6chpt}), which matches experimental data, it
is this latter normalization which corresponds to the JM/CPS treatment
of the scalar channel.  The analogue of the JM/CPS ansatz, in the case
of the vector correlator, thus overestimates the spectral function at
the $\rho$ peak by a factor of $5.1$.

%% NOTE: the difference between 1.28, not 1.24 are the small
%% loop contributions in addition to the LEC contribution.  It is the
%% whole thing that matches well to experiment and hence the whole
%% thing that is the analogue of the JM/CPS treatment. 

The source of this problem is not difficult to identify and, in fact,
turns out to be that the crucial assumption that the spectral function
can be taken to be completely resonance dominated, even near
threshold, is incorrect.  This is most easily seen from the
perspective of {\cpt}.  Indeed, it is known that, when one eliminates
resonance degrees of freedom from a general, extended effective
Lagrangian, producing in the process the usual effective chiral
Lagrangian, ${\cal L}_{eff}^{\cpt}$, relevant to the low-lying
Goldstone boson degrees of freedom alone, the effect of the resonances
present in the original theory is to produce contributions to the
LEC's appearing in ${\cal L}_{eff}^{\cpt}$\onlinecite{EGPdR,Detal}.
There are two important observations about the nature of these
contributions which are of relevance to the present discussion.  The
first is that the resonances do not contribute to the lowest-order
(${\cal O}(q^2)$) LEC's of ${\cal L}_{eff}^{\cpt}$; instead, the
leading (in the chiral expansion) contributions are to the ${\cal
O}(q^4)$ LEC's, $L_k^r(\mu )$ (where $\mu$ is the \cpt\
renormalization scale, and we adhere throughout to the notation of
Gasser and Leutwyler\cite{gl85}).  The second is the phenomenological
observation that, if one takes $\mu\sim m_\rho$, the resonance
contributions essentially saturate the $L_k^r(\mu )$\cite{EGPdR,Detal}
(see, for example, Table 2.1 of Ref.~\cite{BKM95}, for a comparison
with recent experimental determinations of the LEC's).  An immediate
consequence of the first observation is that the correct normalization
for the resonance contributions to quantities like $\rho_{33}$, or
{\me}, near threshold, cannot be that coming from the tree-level
(${\cal O}(q^2)$) \cpt\ contributions, since such contributions are
associated with the Goldstone boson degrees of freedom alone, and
contain no resonance contributions whatsoever.  Similarly, normalizing
to the full threshold value, as obtained, for example, from
experiment, would also be incorrect, since this full value necessarily
contains both tree-level and leading non-analytic contributions,
neither of which can be associated with the resonance degrees of
freedom, in addition to the ${\cal O}(q^4)$ LEC contributions which
{\it do} contain resonance contributions.  Fortunately, the second
observation provides us with an obvious alternative for normalizing
resonance contributions near threshold.  We propose, therefore, to
accept the phenomenological observation above as a general one and
identify resonance effects in near-threshold observables with those
contributions to the 1-loop expressions for these observables
involving the appropriate ${\cal O}(q^4)$ LEC's, $\{ L_k^r\}$,
evaluated at a scale $\mu\sim m_\rho$.  Such an identification,
however, requires that the LEC is dominated by the appropriate
resonance, as is the case for the vector ($L_9$) and scalar ($L_5$)
channels, but not for the pseudo-scalar channel. This prescription,
like that of BPR and JM/CPS, represents a means of using information
solely from the near-threshold region (in this case, obtainable from a
knowledge of the chiral expansion of the spectral function) to
normalize the spectral function in the resonance region.  However, we
will show below that, in contrast to the analogue of the BPR and
JM/CPS ans\"atze which was in error by a factor of $\sim 5$ at the
$\rho$ peak, the new prescription normalizes the peak accurate to
within a few percent.  Based on the success of the prescription in
this channel, we will then apply it to a re-analysis of the JM/CPS
extraction of $m_s$ involving the correlator of the divergences of the
vector current.

Let us return, then, to the spectral function $\rho_{33}$.  According
to the discussion above, the $\rho$ meson contributions to
$\rho_{33}$, near threshold, can be obtained by taking just that term
in Eq.~(\ref{eq:6chpt}) proportional to $L_9^r$, evaluated at a scale
$\mu\sim m_\rho$.  The only change in the above
analysis is then a re-scaling of $\rho^{trial}_{33}$ in Eq.~(\ref{eq:6peak})
by a factor of $0.24$, the value of the ${\cal O}(q^4)$
LEC contribution in Eq.~(\ref{eq:6chpt}) at $\mu =m_\rho$.
This leads to a prediction for the spectral function at the $\rho$
peak of
\be
\label{eq:6LEC}
\rho_{33}^{LEC}(m_\rho^2)=0.067
\ee
in good agreement with the experimental value given in Eq.~(\ref{eq:6rho}).

Let us stress that the precise numerical aspects
of the prescription above, 
namely the supposition that the normalization at resonance peak of 
resonance contributions to the hadronic spectral function
can be obtained by evaluating the relevant
${\cal O}(q^4)$ LEC contributions appearing in near-threshold
\cpt\ expressions, {\it at a scale $\mu\sim m_\rho$}, is one
that is purely phenomenologically motivated \onlinecite{EGPdR,Detal}.  
While highly successful in the case of the vector channel, it has
not been tested outside this channel.  The
fact that resonance contributions begin only at ${\cal O}(q^4)$ in the
chiral expansion, and hence that resonances do not contribute to
either lowest-order tree-level or leading non-analytic terms in the
\cpt\ expansions of the relevant spectral functions, however, clearly
indicates, independent of the numerical reliability of this
prescription, the unsuitability of normalizing the resonance peaks by
associating the full \cpt\ or experimental values near threshold with
resonance effects.  Moreover, so long as the spectral functions of
interest have even reasonably normal chiral expansions, with the
dominant contributions near threshold coming from the lowest order
tree-level contributions, we can conclude that the standard method of
normalization will produce values for these spectral functions in the
resonance region that are overestimated by a significant numerical
factor.  
%Finally, with regard to the question of the numerical
%accuracy, we re-iterate that, as shown above, the prescription works
%very well in the one channel where it can be tested explicitly.

At this stage we should also mention that Stern and collaborators have
suggested that the normalization at threshold could actually be much
larger than that given by leading order {\cpt}, as is expected in
``generalized {\cpt}''\cite{Stern93}. They then argue that, in that 
case, the quark masses would be even larger. Our observations are also
relevant in this case: we again stress that, since the sum rules we
consider are dominated by the resonance region, 
threshold normalization will only provide useful input if one can
disentangle the contributions to threshold amplitudes associated
with resonances from those associated with the Goldstone boson
degrees of freedom.

\section{Re-analysis of the JM/CPS extraction of \strangem}
\label{s:s_JM}

In this section we will employ the prescription proposed above to a 
re-analysis of the JM/CPS extractions of $m_s$\cite{CPS96,jm95}.  
Such a re-analysis is possible in this case because the 1-loop
\cpt\ expression for the relevant scalar form factor is known
\onlinecite{gl85ff}.  To introduce notation, we briefly review the
analysis of Ref.~\cite{jm95} (that of Ref.~\cite{CPS96,cdps} is
similar and need not be discussed separately). These analyses involve
a standard QCD sum rule treatment of the correlation function
\bea
\label{eq:mscorrelator}
\Psi (q^2) &=& {\rm i}\int d^4x\ e^{i\, q.x}\langle 0\vert T\{ \partial^\mu
V_\mu (x) \partial^\nu {V^\dagger}_\nu (0)\}\vert 0\rangle \nonumber \\
&=& (m_s -m_u)^2\, {\rm i}\int d^4x\ e^{i\, q.x}\langle 0\vert T\{ S(x) 
S^\dagger (0)\}\vert 0\rangle ,
\eea
where $V_\mu (x)$ is the strangeness-changing vector current and
$S(x)$ the corresponding strangeness-changing scalar current.  The
correlator of scalar currents is evaluated using the operator product
expansion (OPE). All terms on this side of the sum rule are
proportional to $(m_s -m_u)^2$, and the full $\alpha_s^3$ pQCD result
is known for the predominant contribution $\Psi^{''}_0$
\cite{CPS96}. The hadronic spectral function in the phenomenological
side is again taken to be a sum-of-resonances modulation of the
spectral function relevant to the $K\pi$ intermediate state near
threshold.

%One of the strong points in favor of the sum rule analyses is that the
%ratio $m_s / (m_u + m_d) \sim 180/12 = 15$ comes out in agreement with
%the predictions of \cpt. If our arguments concerning the $O(\alpha^3)$
%terms in the pQCD results, decreasing the normalization $A$ in the
%hadronic $\rho_5$, and increasing $s_0$ are correct, then the estimate
%of \mbar\ could easily be half the quoted value.  Concomitantly, to
%keep the ratios predicted by \cpt\ intact would require a similar
%reduction in the value of $m_s$. To understand how this could happen
%we now turn to the sum rule analyses of Refs.~\cite{jm95,cdps} for
%$m_s$.

JM/CPS write the $K\pi$ contribution to the physical spectral function as 
\be
\label{eq:62}
\rho_{K\pi}(s)={\frac{3}{32\pi^2 s}}\theta (s-s_+) \sqrt{(s-s_+)(s-s_-)}
\, \vert d(s)\vert^2
\ee
where $s_\pm =(m_K\pm m_\pi )^2$, and $d(s)$ is the strangeness-changing
scalar form factor, measured in $K_{\ell 3}$ for $m_\ell^2\leq s\leq s_-$,
\be
\label{eq:63}
d(s)\equiv (m_K^2 -m_\pi^2)f_0(s)=(m_K^2 -m_\pi^2)f_+(s) + s f_-(s)
\ee
with $f_\pm (s)$ the usual form factors defined by
\be
\label{eq:64}
\langle \pi^0(p^\prime )\vert \bar{s} \gamma_\mu u\vert K^+(p)\rangle =
{\frac{1}{\sqrt{2}}}\left[ (p^\prime +p)_\mu f_+(s) + (p-p^\prime )_\mu
f_-(s)\right]\ .
\ee
In their analysis, JM/CPS employ the following resonance-modulation ansatz
for the spectral function:
\be
\label{eq:65}
\rho_{hadronic} (s) = {\frac{3}{32\pi^2 s}}\sqrt{(s-s_+)(s-s_-)}\, \vert d(s_+)\vert^2\,
F(s)
\ee
where
\be
\label{eq:66}
F(s)=\frac{\sum_n c_n^{BW}(s)}{\sum_n c_n^{BW}(s_+)}
\ee
with
\be
\label{eq:67}
c_n^{BW}(s)=\frac{f_n^2 m_n^5\Gamma_n}{(m_n-s)^2+m_n^2\Gamma_n^2(s)}\ .
\ee
In Eqs.~(\ref{eq:65})-(\ref{eq:67}),
$s_+$ is the continuum $K\pi$ threshold, and $f_n$, $m_n$, and
$\Gamma_n$ are the decay constant, mass and width of the {\it nth}
scalar resonance, $\Gamma_n(s)$ being the usual s-dependent width
given in \cite{jm95}.  The s-dependence of the width factor occurring
in the numerator of the Breit-Wigner resonance forms has already been
factored out explicitly in writing Eq.~(\ref{eq:65}).  The sum in
Eq.~(\ref{eq:66}) is taken to run over two resonances (the
$K_0^*(1430)$ and $K_0^*(1950)$), and the duality point, $s_0$, of QCD
sum rules (describing the point beyond which the physical spectral
function is to be modeled by its perturbative expression) is fixed by
a stability analysis.  Note that the normalization procedure above
assumes that the physical spectral function is completely saturated by
resonance contributions near threshold.  The threshold value of the
scalar form factor, $d(s_+) = 0.33\pm0.02 \GeV^2$, is obtained using 
Omnes representation with experimental $K\pi$ phase shifts as input. 
This result is, moreover, shown to be consistent with that of \cpt\ to
l-loop, which can be obtained from the expression for $f_0(s)$ given
by Gasser and Leutwyler\cite{gl85ff} ($d_0^{\chi PT} (s_+)=0.35\
\GeV^2$).  Lastly, the master equation used for extracting $m_s$ is \cite{jm95}
\be
\label{eq:JMmaster}
u^3 \hat{\Psi}^{''}_{OPE} \ = \ \int_0^{s_0} e^{-s/u} \rho_{hadronic}  \ ds + 
                                \int_{s_0}^\infty e^{-s/u} \rho_{pQCD} \ ds
\ee
where both $\hat{\Psi}^{''}_{OPE}$ and $\rho_{pQCD}$ are proportional to 
$(m_s - m_u )^2$. 

The first of the three issues raised by us, namely the reliability of
pQCD has already been discussed in Section~\ref{s:sPQCD}. We
agree with JM/CPS that in this channel the effect of the neglected
$\alpha_s^4$ and higher contributions could, at best, lower estimates
of $m_s$ by $\sim 5\%$.  The remaining two issues, the value of $s_0$ 
and the normalization of the hadronic spectral function are far more 
serious, as we now explain. 

To elucidate the role of $s_0$ in the JM/CPS analysis we plot, in
Fig.~\ref{f_JMrho}, both the model JM hadronic spectral function (for
$s< s_0$) and the pQCD version of the spectral function (for $s>
s_0$).  We have used the JM values corresponding to the preferred
solution, $i.e.$, $s_0=6.0\ \GeV^2$, $\Lambda_{QCD}^{(3)}=380 \MeV$ and
$m_s=189 \MeV$.  The plot shows very clearly that the ansatz for
$\rho_{hadronic}$ is, at best, valid only for $s \le 4.0 \GeV^2$.
Furthermore, as evident from Eqs.~(\ref{eq:65})-(\ref{eq:67}),
$\rho_{hadronic}$ goes to a constant at large $s$, whereas
$\rho_{pQCD}$ grows linearly (with logarithmic corrections).  For this
reason there is a large discontinuity between $\rho_{hadronic}$ and
$\rho_{pQCD}$ even for $s$ as low as $4 \GeV^2$.  The only way that
$\rho_{hadronic}$ constructed from the $K \pi$ channel can satisfy
duality is if there is a piling up of higher resonances, and these
have to have large amplitudes (as we illustrated in section
\ref{s:sguess} for the pseudoscalar channel).  We contend that $s_0$
should only be chosen in the range where $\rho_{hadronic}$ is known
reliably.  However, for $s_0 \le 4.0 \GeV^2$, and using the JM/CPS
ansatz for $\rho_{hadronic}$, we have not been able to find a result
for $m_s$ that is stable under variations of the Borel parameter $u$.
It was precisely this lack of stability that forced JM/CPS to choose a
larger $s_0$. Such a choice, we contend, is not reasonable as
$\rho_{hadronic} \ll \rho_{pQCD}$ over the range $ 3 < s < 6 \GeV^2$,
$i.e.$, duality is badly violated over this whole range.

\iffigure
\begin{figure}[t]   %7
\hbox{\hskip15bp\epsfxsize=0.9\hsize \epsfbox {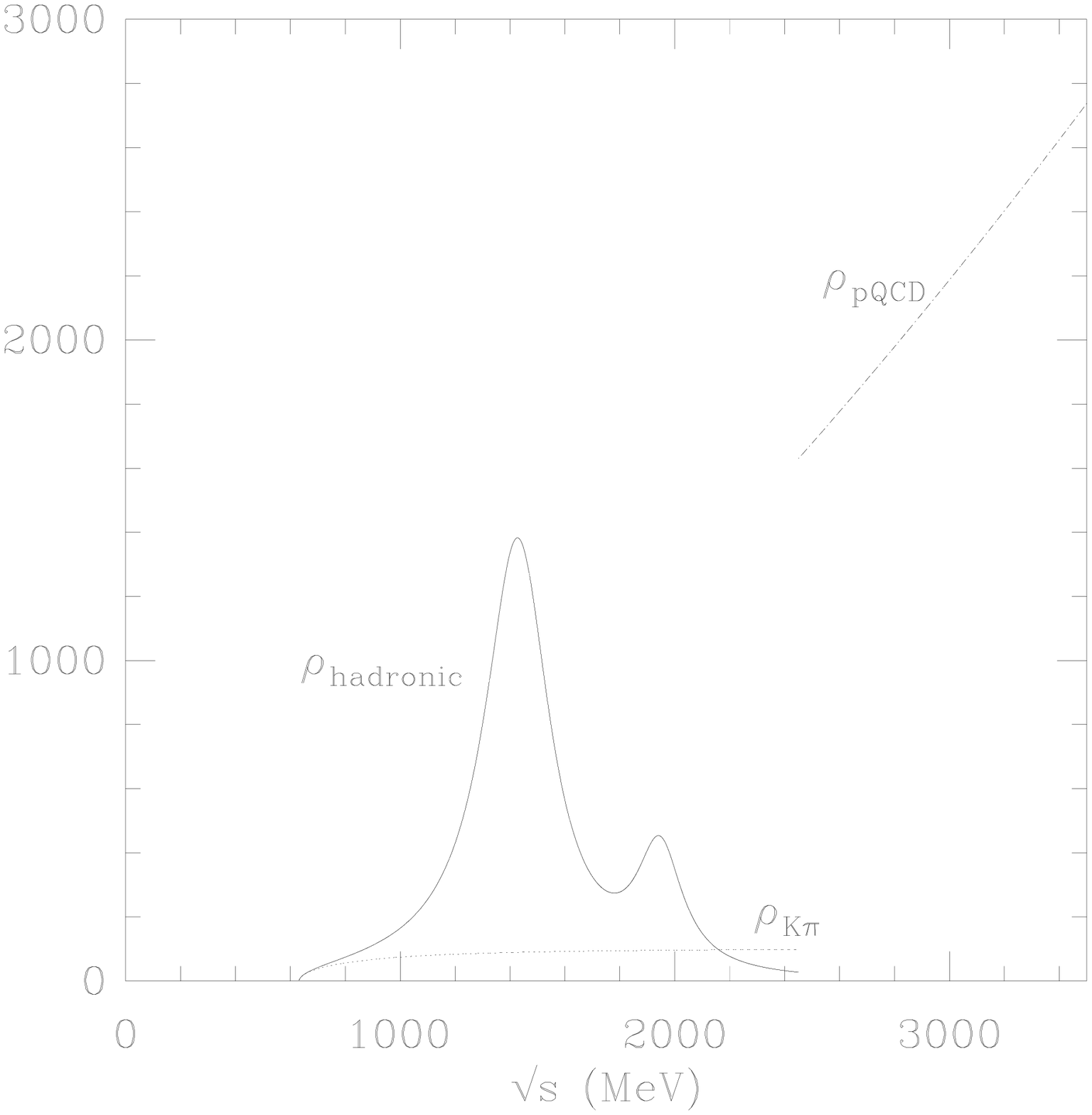}}
\figcaption{Plots of the spectral functions $\rho_{pQCD}$ and
$\rho_{hadronic}$ used by JM/CPS \cite{jm95,CPS96}. The scale of
matching between the pQCD and hadronic solution is $s_0 = 6.0 \GeV^2$.
To highlight the fact that the $\rho_{hadronic}$ is dominated by the
resonance contribution we also show $\rho_{K \pi}$, $i.e.$
$\rho_{hadronic}$ without the Breit-Wigner modulation factor. For
convenience we plot $\rho*10^5/(m_s-m_u)^2$, so the units along the
$y$ axis are $\GeV^2$. The values, $m_s=189$ MeV and $m_u =5$ MeV have 
been taken from Ref.~\cite{jm95}.}
\vskip \baselineskip
\label{f_JMrho}
\end{figure}
\fi

Lastly, we turn to the quantity, $d(s_+)$, which sets the overall
normalization of the resonance contributions in Eq.~(\ref{eq:65}).
This quantity is crucial in the JM/CPS analysis since, as noted by JM,
the extracted value of $m_s$ scales directly with $d(s_+)$.  The
problem is that, just as for the light quark case, the spectral
integral appearing on the phenomenological side of the sum rule is
dominated, not by the near-threshold region, but by resonance
contributions.  The ansatz (\ref{eq:65})-(\ref{eq:67}) for the
spectral function, however, is designed only to produce the correct
overall normalization at $K\pi$ threshold.  From our discussion above
of the analogous treatment of the vector current correlator, it is
clear that such an ansatz will overestimate the resonance
contributions near threshold, and hence almost certainly significantly
overestimate the spectral function in the resonance region.  To
correct this problem we need to properly rescale the JM/CPS ansatz at
threshold.  We do so on the basis of the proposal above, $i.e.$, we
assume that in the scalar channel, just as in the vector channel, the
${\cal O}(q^4)$ LEC's, evaluated at a scale $\mu\sim m_\rho$, give the
correct normalization of the scalar resonance contributions at
threshold.  It is easy to implement this revised normalization of
$\rho_{hadronic}$ because, not only is the 1-loop \cpt\ expression for
$d(s)$ known\cite{gl85ff}, but, in addition, Jamin and M\"unz have
demonstrated explicitly the accuracy of this expression for
$d(s_+)$\cite{jm95}.

Let us write the 1-loop \cpt\ expression for $d(s_+)$ in the form
\be
\label{eq:dchiral}
d_{\cpt}(s_+) = d_{tree}(s_+) + d_{res}(s_+,\mu )
+ d_{loop}(s_+,\mu )
\ee
where $ d_{tree}(s_+)$ is the leading, ${\cal O}(q^2)$ tree-level
contribution, $ d_{res}(s_+,\mu )$ contains the ${\cal O}(q^4)$ LEC
contributions, and $ d_{loop}(s_+,\mu )$ contains the contributions
associated with 1-loop graphs generated from the ${\cal O}(q^2)$
part of the effective chiral Lagrangian.  The latter two terms are
separately scale-dependent.  According to the prescription introduced
above, resonance contributions to $d(s_+)$ are to be identified with
$ d_{res}(s_+,m_\rho )$.  Resonance contributions to
$\vert d(s_+)\vert^2$, consistent to 1-loop order, are thus given by
\be 
\label{eq:dres}
\vert d(s_+)\vert^2_{res}\simeq 2\, d_{tree}(s_+)\, d_{res}(s_+, m_\rho )\ .
\ee
Using\cite{gl85ff}
\bea
\label{eq:dnum}
d_{tree}(s_+)&= (m_K^2-m_\pi^2)= 0.22 {\rm\ GeV}^2\nonumber \\
d_{res}(s_+,m_\rho )&= 4s_+ (m_K^2-m_\pi^2) L^r_5(m_\rho )/f_\pi^2\ ,
\eea
with $ L^r_5(m_\rho )= 0.0014\pm 0.0005$, we find that $\vert
d(s_+)\vert^2_{res}\sim 0.23 \vert d(s_+)\vert^2$.  With no other changes
to the JM/CPS analysis than the corresponding re-scaling of the
continuum spectral function, the value of $m_s$ would thus be lowered
by almost exactly a factor of $2$.  However, as discussed above,
there are problems of consistency with using the JM/CPS ansatz for
the spectral function with values of $s_0$ as large as $6$ GeV$^2$.

In light of the above corrections, the question before us is whether
it is possible to get a stable estimate of $m_s$ by repeating the
JM/CPS analysis with $s_0 \approx 4 \GeV^2$ and an overall
normalization of $\rho_{hadronic}$ of $A \approx 0.25$.  To answer
this question we have varied $\Lambda_{QCD}^{(3)}$ in the range $200-450$
MeV, the relative strength $f_2/f_1$ of the two Breit-Wigner
resonances in the modulating factor over $0.2-1$, and $A$ over the
range $0.2 - 1$.  Despite this, we have failed to find a solution that
is stable under variations in the Borel scale $u$. The cause of this
failure is the ansatz for $\rho_{hadronic}$, and the small range of
$s$ over which it can be evaluated. It is our contention that reliable
results for $m_s$ using sum-rules can only be obtained if
$\rho_{hadronic}$ is determined to high precision over a sufficiently
large range of scales, say from $s_{th}$ to $8 \GeV^2$. If $s_0$ is
``small'' then limitations of operator product expansion, convergence
of perturbation theory at small $s$, and details of resonances
contributions make it difficult to test the reliability of the
results.

%% Incorporating such a correction, the JM estimate of $m_s$ will be
%% lowered by a factor of $\sqrt{0.23}=0.48$, yielding
%% \begin{equation}
%% m_s=91\pm 15\pm 32\ {\rm MeV}
%% \end{equation}
%% where the first error is that of the JM analysis and the second is
%% that associated with the error in the phenomenological determination
%% of $L^r_5(m_\rho )$, which probably also serves as a reasonably
%% conservative estimate of the systematic error associated with the
%% phenomenological nature of the prescription employed in normalizing the
%% hadronic spectral function.  Combining this result with the
%% constraints on the ratio, $r$, obtained from {\cpt}\cite{Leutwyler96},
%% we obtain
%% \begin{equation}
%% m_u+m_d=7.5\pm 3 \ {\rm MeV}\ ,
%% \end{equation}
%% again compatible with the lattice result.  Note that additional
%% corrections for $O(\alpha_s^3)$ effects, or an increase in $s_0$,
%% could easily further lower the estimate for $m_s$, so that
%% compatibility even with the much lower central value of the unquenched
%% lattice analysis is possible.  {\bf Note that I've re-instated the
%% numerical values for $m_s$ and \summ\ .  I've added additional
%% comments suggesting why I think the re-analysis is reasonably
%% plausible and, since, in any case it is vastly more plausible than
%% previous analyses, we should quote the numbers explicitly!!!! (in my
%% opinion).  I have however slightly altered the discussion of errors.}

For completeness we should also mention the alternate JM determination
of $m_s$ via an analysis of the analogous strangeness-changing axial
correlator. Their result in this case is $m_s(\MSbar,2\GeV) = 91
\MeV$, significantly smaller than that obtained from the scalar
channel via the treatment of the vector current correlator.  They,
however, consider this analysis incomplete because it employs, for the
normalization of the continuum spectral function at threshold, the
leading-order, tree-level {\cpt} result.  They contend,
based on the
expectation that the full normalization will, as in the scalar channel, 
significantly exceed that given by
tree-level \cpt\ ($d(s_+) = 1.5 d_{tree}(s_+)$ for the scalar channel),
that the true
normalization will likely be significantly larger .  If true,
this would mean that $m_s$ would be correspondingly increased.  They
thus expect their two analyses to become consistent once they employ a
normalization at threshold corresponding 
to the 1-loop expression for the continuum
spectral function in the pseudoscalar channel.  Our
contention is that, in fact, the ``correct'' normalization is given, not by
the full threshold spectral function, but rather by
the appropriate ${\cal O}(q^4)$ 
LEC contributions to the 1-loop result, and that it should hence be
significantly {\it smaller} than that corresponding to the
tree-level result.  Further progress on this issue, and that of the
consistency of the two different extractions for $m_s$, will be possible only
once the 1-loop expression for $<0|\partial_\mu A^\mu
|K\pi\pi >$ is known \cite{cwkrm}.  A re-analysis of the BPR FESR treatment
of $m_u+m_d$ is similarly stymied by the absence of 1-loop
expressions for the matrix elements $<0\vert \partial^\mu A_\mu^{(\pm )}
\vert 3\pi >$, and by the lack of association of the $L_i$ involved with 
just the pseudoscalar resonances.

In the past, of course, the
agreement of the ratio $r\sim 2(180)/12=30$ obtained from the
different sum rule analyses with that ($24.4\pm 1.5$) obtained from
{\cpt}\cite{Leutwyler96} has been taken to provide {\it a posteriori}
support for the validity of the sum rule treatments.  Our contention
is that a self-consistent sum-rule analysis would yield estimates of
both $m_s$ and \summ\ that are lower by a factor of $\sim 2$, thus
maintaining the consistency with the \cpt\ value of $r$.

\section{Conclusions}
\label{s:s_conclusions}

We have shown that the ability to make reliable extractions of $m_s$
and $m_u + m_d$ using sum rule analyses rests on three key features of
these analyses: the degree of reliability of pQCD, a knowledge of the
scale, $s_0$, at which quark-hadron duality becomes valid, and an ability to
construct hadronic spectral functions which are correctly
normalized in the resonance region, even in channels where experimental
data on the relevant decay constants is not available.

We find that, in the relevant pQCD expressions, 
the $\alpha_s, \alpha_s^2, \ldots$ corrections are large both
in the scalar and pseudoscalar channels.  Including reasonable
estimates for the unknown higher order terms lowers the sum rule
estimates of quark masses.  The largest effect is in the
extraction of $\mbar$, which we estimate would be lowered by $\approx
20\%$ compared to the value quoted by BPR \cite{BPR95}.  The
correction in the case of $m_s$ extracted from the scalar channel is
roughly $5\%$, and this has been accounted for by JM/CPS.

Second, the estimates obtained for the quark masses are potentially
very sensitive to the choice of $s_0$. We have illustrated this through an
analysis of rigorous lower bounds and
the use of a variety of plausible spectral functions in
the case of $\mbar$. Current sum rule analyses are forced to choose
low values of $s_0$ due to lack of experimental information. 
The FESR extraction of {\mbar}, for example, is based on rather low values
of $s_0 \le 3 \GeV^2$, so no tests of the stability of the estimates
under variations of $s_0$ can be made.  In the case of JM/CPS analysis
of $m_s$, the value chosen, $s_0=6.0 \GeV^2$, is artificially
large.  This choice arises from an attempt 
to achieve stability of the Borel transformed sum rule with respect to the
Borel parameter $u$.  Since, however, the phenomenological ansatz for the
spectral function breaks down for $s \gsim 4.0 \GeV^2$, it
is clear that such a choice of $s_0$ is not physical. For reasonable
choices of $s_0$ we are also not able to find a solution that is stable
with respect to variations in $u$. We, therefore, conclude that no
reliable estimates of $m_s$ can be made unless $\rho_{hadronic}$ is
known accurately over a significantly larger range of $s$.

Third, we have shown that the method employed in previous analyses for
fixing the overall normalization of the resonance-modulated model
spectral functions leads to significant overestimates of the continuum
contributions to the relevant spectral integrals and hence to
significant overestimates of the quark masses.  The source of this
problem is the fact that normalizing the resonance-modulated ansatz
(see Eqs.~(\ref{eq:65})-(\ref{eq:67})) to either the experimental
value or to the \cpt\ value for the spectral function in the
near-threshold region results in the inclusion of near-threshold
contributions of the Goldstone-boson degrees of freedom in addition to
the desired resonance contributions. Overestimating the size of the
resonance tail in this manner, of course, leads to a corresponding
overestimate of the resonance contributions at resonance peak.  Since
it is the resonance peak region, and not the threshold region, which
dominates the phenomenological side of the sum rules, the conventional
procedure produces significant overestimates of the quark masses.  In
the case of the vector current correlator, where the normalization of
the spectral function at the $\rho$ peak is known experimentally, we
have shown that the magnitude of this overestimate is large: the
conventional method of normalization produces a spectral function
which, at the $\rho$ peak, is a factor $ 4.1 - 5.1$ larger than that
given by experiment.  We have explained, based on an understanding of
the manner in which resonance effects manifest themselves in {\cpt}
why the conventional method of normalization cannot be correct, and
have proposed an alternate
phenomenological prescription for normalizing the spectral function,
designed to provide estimates which are reliable not so much in the
threshold region as in the resonance region relevant to the sum rule
quark mass extractions.  We verify that this prescription reproduces 
the experimental result for the vector ($\rho$) channel. 
This method is straightforward to apply to 
the scalar channel as the 1-loop ($O(q^4)$) \cpt\ corrections are known, 
and the revised estimate for the normalization could reduce 
the estimate of $m_s$ by as much as a factor of
$\sim 2$ over the values found in previous analyses.
We argue that a similar overestimate of the
normalization will exist in the pseudoscalar channels,
though we are unable to estimate its magnitude at present.

The bottom line is that unless the hadronic spectral function is known
accurately over a large range of scales, say up to $s = 8 \GeV^2$,
reliable extraction of quark masses from sum-rules considered is not
possible. 
\ifFINAL
\else
For completeness, we have also presented a reanalysis of the
possible systematic errors in lattice QCD estimates. 
\fi
Even though the
lattice QCD estimates have their share of statistical and systematic
errors~\cite{Mq97TBRG}, we claim that at present they represent the most reliable
means of estimating the quark masses.  Our estimates of the systematic
errors in sum rule analysis suggest that revised sum rules estimates
could easily be smaller by a factor of two, in which case these would
be consistent with the small values obtained from lattice QCD.

\section*{Acknowledgements}
We are very grateful to J. Bijnens, J.~Padres, and E.~de Rafael, to
M.~Jamin, and to D.~Pirjol for communicating details of their analyses
to us and for discussions. We also thank S.~Godfrey and H.~Blundell
for communicating previously unpublished results of their model of the
meson spectrum and decays.  K.~Maltman thanks the T5 Group, Los Alamos
National Laboratory, and the Special Research Center for the Subatomic
Structure of Matter of the University of Adelaide for hospitality
during the course of this work and also acknowledges the ongoing
financial support of the Natural Sciences and Engineering Research
Council of Canada.

%%%%%%%%%%%%%%%%%%%%%%%%%%%%%%%%%%%%%%%%%
%%%%%%%%%%%%%%%%%%%%%%%%%%%%%%%%%%%%%%%%%
%%%%%%%%%%%%%%%%%%%%%%%%%%%%%%%%%%%%%%%%%

%

\end{document}

%% file: rgmac.latex
\newskip\defaultbaselineskip\defaultbaselineskip=12pt

\def\GeV{\mathord{\rm \;GeV}}
\def\MeV{\mathord{\rm \;MeV}}

\def\bar{\overline}

\def\eqinf{\mathrel{\raise2pt\hbox to 24pt{\raise -8pt\hbox{$t \to \infty$}\hss{$=$}}}}

\def\gsim{\mathrel{\raise2pt\hbox to 8pt{\raise -5pt\hbox{$\sim$}\hss{$>$}}}}
\def\rsim{\mathrel{\raise2pt\hbox to 8pt{\raise -5pt\hbox{$\sim$}\hss{$>$}}}}
\def\lsim{\mathrel{\raise2pt\hbox to 8pt{\raise -5pt\hbox{$\sim$}\hss{$<$}}}}
\def\ssqr#1#2{{\vbox{\hrule height.#2pt
      \hbox{\vrule width.#2pt height#1pt \kern#1pt\vrule width.#2pt}
      \hrule height.#2pt}\kern-.#2pt}}

\def\Dsl{\,\raise.15ex\hbox{$/$}\mkern-13.5mu D} %this one can be subscripted
\def\dsl{\raise.15ex\hbox{$/$}\kern-.57em\hbox{$\partial$}}

\ifx\href\undefined\def\href#1#2{{#2}}\fi
\def\spireshome{http://www.slac.stanford.edu/cgi-bin/spiface/find/hep/www?FORMAT=WWW&}
{\catcode`\%=12
\xdef\spiresjournal#1#2#3{\noexpand\protect\noexpand\href{\spireshome
                          rawcmd=find+journal+#1%2C+#2%2C+#3}}
\xdef\spireseprint#1#2{\noexpand\protect\noexpand\href{\spireshome rawcmd=find+eprint+#1%2F#2}}
\xdef\spiresreport#1{\noexpand\protect\noexpand\href{\spireshome rawcmd=find+rept+#1}}
\xdef\spireskey#1{\noexpand\protect\noexpand\href{\spireshome key=#1}}
}
\def\eprint#1#2{\spireseprint{#1}{#2}{#1/#2}}

\def\nohref{}

\def\putpaper{\edef\refpage{\the\count0}%
              \def\nohref{}%
              {\def\ {+}\def\nohref##1{}\edef\temp{\noexpand\spiresjournal
               {\journalname}{\volume}{\refpage}}\expandafter}\temp
               {\sfcode`\.=1000{\journalname} {\bf \volume} (\refyear)
                \refpage}\egroup}
\def\putpage{\edef\refpage{\the\count0}%
              \def\nohref{}%
              {\def\ {+}\def\nohref##1{}\edef\temp{\noexpand\spiresjournal
               {\journalname}{\volume}{\refpage}}\expandafter}\temp
              {\refpage}\egroup}
\def\dojournal#1#2 (#3){\def\journalname{#1}\def\volume{#2}\def\refyear
                        {#3}\afterassignment\putpaper\bgroup\count0=}
\def\morepage{\afterassignment\putpage\bgroup\count0=}
\def\supresslink{\def\spiresjournal##1##2##3{}}

\def\APNY#1{\dojournal{Ann.\ Phys.\ \nohref{(N.\ Y.)}}{#1}}

\def\IJMPE#1{\dojournal{Int.\ J.\ Mod.\ Phys.}{E#1}}

\def\MPLA#1{\dojournal{Mod.\ Phys.\ Lett.}{A#1}}

\def\NPB#1{\dojournal{Nucl.\ Phys.}{B#1}}
\def\NPBPS#1{\dojournal{Nucl.\ Phys.\ \nohref(Proc.\ Suppl.\nohref)}{\nohref B#1}}

\def\PRD#1{\dojournal{Phys.\ Rev.}{D#1}}

\def\PLB#1{\dojournal{Phys.\ Lett.}{#1B}}

\def\ZPC#1{\dojournal{Z.\ Phys.}{C#1}}

\def\etal{{\it et al.}}

%% file: t_fitcomp.tex
%% t_norms.tex    
%% Normalizations for the 4 cases of spectral functions
%% 
\setlength{\tabcolsep}{3pt}
$$
\begin{tabular}{|l|c|c|c|c|c|c|c|}
\hline
$         $&$ s_d(\GeV^2) $&$\summ(\MeV)$&$ c_1    $&$ c_2          $&$ c_3    $&$ c_4    $&$ A    $\cr
\hline 
$ Case\ 1 $&$ 3.0         $&$ 12.0  $&$ 1      $&$ -0.23+0.65i  $&$  -     $&$ -      $&$ 1.0  $\cr
$ Case\ 2 $&$ 5.7         $&$  9.0  $&$ 1      $&$ 1.0          $&$ 2.3    $&$ -      $&$ 1.0  $\cr
$ Case\ 3 $&$ 8.0         $&$  8.0  $&$ 1      $&$ 1.2          $&$ 5.0    $&$ 6.5    $&$ 0.7  $\cr
$ Case\ 4 $&$ 10.0        $&$  6.0  $&$ 1      $&$ 0.8          $&$ 2.0    $&$ 3.68   $&$ 0.5  $\cr
\hline
\end{tabular}
$$

%% file: t_fitssd10.tex
%% t_norms.tex    
%% Normalizations for the s_d=10, mbar = 6 solution 
%% 
\setlength{\tabcolsep}{3pt}
$$
\begin{tabular}{|l|c|c|c|c|c|}
\hline
$         $&$ c_1    $&$ c_2   $&$ c_3    $&$ c_4    $&$ A    $\cr
\hline 
$ Case\ 1 $&$ 1      $&$ 0.8   $&$ 2.0    $&$ 3.680  $&$ 0.5  $\cr
$ Case\ 2 $&$ 1      $&$ 0.6   $&$ 2.0    $&$ 5.160  $&$ 0.4  $\cr
$ Case\ 3 $&$ 1      $&$ 1.0   $&$ 4.0    $&$ 11.75  $&$ 0.3  $\cr

\hline
\end{tabular}
$$